\documentclass[prb,aps,twocolumn,superscriptaddress,showpacs,amsmath,amssymb]{revtex4-1}

\usepackage{graphicx}
\usepackage{epstopdf,color}
\usepackage{amsfonts,dsfont}
\usepackage{bm,braket,array}
\usepackage[enableskew]{youngtab}
\usepackage{pst-all}
\usepackage[off]{auto-pst-pdf}

\usepackage{epsfig,wrapfig,subfigure,rotating,wasysym} 
\usepackage{tikz}
\usetikzlibrary{calc,arrows,decorations.pathreplacing,decorations.markings}

\usepackage{dcolumn}
\usepackage{bm}
\allowdisplaybreaks 
\usepackage{braket}
\usepackage{ulem} 
\usepackage{hyperref}
\hypersetup{colorlinks=true, linkcolor=blue, citecolor=blue, filecolor=blue, urlcolor=blue, pdftitle=, pdfauthor=, pdfsubject=, pdfkeywords=}

\newcommand{\bs}[1]{\boldsymbol{#1}}

\newcommand{\comm}[2]{\left[#1,#2\right]}

\newcommand{\half}{$\frac{1}{2}$ }
\renewcommand{\i}{\text{i}}

\newcommand{\x}{\text{x}}
\newcommand{\y}{\text{y}}
\newcommand{\z}{\text{z}}

\newcommand{\up}{\uparrow}
\newcommand{\dw}{\downarrow}

\newcommand{\vac}{\left|0\right\rangle}

\renewcommand{\a}{\alpha}
\renewcommand{\b}{\beta}
\renewcommand{\c}{\gamma}
\newcommand {\ea}{\eta_{\alpha}}
\newcommand {\eb}{\eta_{\beta}}
\newcommand {\ec}{\eta_{\gamma}}

\newcommand {\eab}{\bar\eta_{\alpha}}
\newcommand {\ebb}{\bar\eta_{\beta}}

\newcommand {\bSa}{\bs{S}_{\alpha}}
\newcommand {\bSb}{\bs{S}_{\beta}}
\newcommand {\bSc}{\bs{S}_{\gamma}}

\def\s{\scriptscriptstyle}

\newlength{\ytlength}
\def\ie{{i.e.},\ }

\def\etal{{et al.}}

\begin{document}
\title {Non-Abelian Statistics in one dimension: 
  topological momentum spacings and SU(2) level $k$ fusion rules}

\author{Martin Greiter} \affiliation{Institute for Theoretical Physics, 
  University of W\"{u}rzburg, Am Hubland, D-97074 W\"{u}rzburg, Germany}

\author{F.D.M. Haldane} \affiliation{Department of Physics, Princeton University, Princeton, New Jersey 08544, USA}

\author{Ronny Thomale} \affiliation{Institute for Theoretical Physics, 
  University of W\"{u}rzburg, Am Hubland, D-97074 W\"{u}rzburg, Germany}

\date{\today}


\begin{abstract} 
We use a family of critical spin chain models discovered recently by one of us [M. Greiter, \emph{Mapping of Parent Hamiltonians}, Springer, Berlin/Heidelberg 2011] to propose and elaborate that non-Abelian, SU(2) level $k=2S$ anyon statistics manifests itself in one dimension through topological selection rules for fractional shifts in the spacings of linear momenta, which yield an internal Hilbert space of, in the thermodynamic limit degenerate states.  These shifts constitute the equivalent to the fractional shifts in the relative angular momenta of anyons in two dimensions.  We derive the rules first for Ising anyons, and then generalize them to SU(2) level $k$ anyons.  We establish a one-to-one correspondence between the topological choices for the momentum spacings and the fusion rules of spin \half spinons in the SU(2) level $k$ Wess--Zumino--Witten model, where the internal Hilbert space is spanned by the manifold of allowed fusion trees in the Bratelli diagrams.  Finally, we show that the choices in the fusion trees may be interpreted as the choices between different domain walls between the $2S+1$ possible, degenerate dimer configurations of the spin $S$ chains at the multicritical point.  \end{abstract}


\maketitle

\section{Introduction \label{sec:intro}}
%
The concept of fractional quantization, and, in particular, fractional
statistics~\cite{wilczek90}, is witnessing a renaissance of interest
in recent years.  This is due to possible applications of states
supporting excitations with non-Abelian statistics~\cite{stern10n187}
to the rapidly evolving field of quantum computing and cryptography.
The paradigm for this class is the Pfaffian
state~\cite{moore-91npb362,greiter-92npb567}, which has been proposed
to describe the experimentally observed quantized Hall plateau at
Landau level filling fraction
$\nu=\frac{5}{2}$~\cite{greiter-92npb567}.  The state supports
quasiparticle excitations which possess Majorana fermion states at
zero energy~\cite{read-00prb10267}.  Braiding of these half-vortices
yields non-trivial changes in the occupations of the Majorana fermion
states, and hence renders the exchanges non-commutative or
non-Abelian~\cite{ivanov01prl268,stern-04prb205338}.  Since this
``internal'' state vector is insensitive to local perturbations, it is
preeminently suited for applications as protected qubits in quantum
computation~\cite{kitaev03ap2,nayak-08rmp1083}.  Non-Abelian anyons
are further established in other quantum Hall states including
Read-Rezayi states~\cite{read-99prb8084}, in the non-Abelian phase of
the Kitaev model~\cite{kitaev06ap2}, the Yao--Kivelson and Yao--Lee
models~\cite{yao-07prl247203,yao-11prl087205}, and in the family of
non-Abelian chiral spin liquid (NACSL) states introduced by
two of
us~\cite{greiter-09prl207203,scharfenberger-11prb140404,greiter-14prb165125}. The
latter were subsequently
revisited from a coupled ladder
approach~\cite{meng-15prb241106,lecheminant-17prb140406}, continuum
limit interpolation~\cite{glasser-15njp082001}, as well as from the viewpoint of
Gutzwiller-projected superconductors~\cite{liu-18prb195158} and
topological entanglement entropy~\cite{wildeboer-16prb045125}.

In this article, we propose and elaborate that the possibility of
non-Abelian statistics is not limited to two spacial dimensions, but
exists in certain families of one dimensional spin models as well.  In
particular, we use a family of (multi-) critical spin chain models
discovered recently by one of us~\cite{Greiter11} (and for SU(2) level
2 shortly thereafter independently obtained by Nielsen
\etal~\cite{nielsen-11jsmte11014}), to show that the SU(2) level $k$
anyon statistics of the spinons of these models manifests itself in
one dimension through topological selection rules for the fractional
momentum spacings, which constitute the equivalent to fractional
relative angular momentum for anyons in two dimensions.  In Section
\ref{sec:hs}, we review the fractional momentum spacings of the
spinons excitation in an $1/r^2$ model discovered by Shastry and one
of us, commonly referred to as the Haldane--Shastry
model~\cite{haldane88prl635,shastry88prl639}.  This model constitutes
a lattice 
Wess--Zumino--Witten
model~\cite{wess-71plb95,witten84cmp455,DiFrancescoMathieuSenechal97},
and the spinons obey Abelian half-Fermi
statistics~\cite{haldane91prl937,greiter09prb064409}.  We further
review the formalism of extended Young
tableaux~\cite{greiter-07prl237202,scharfenberger-12jpa455202}, which
provides the single particle momenta of the spinon excitations.  In
Section \ref{sec:na}, we first review the ground state and the
associated parent Hamiltonian of a similar, exactly solvable, critical
spin model for a spin $S=1$ chain~\cite{Greiter11,thomale-12prb195149,
  nielsen-11jsmte11014}.  This model supports spinon excitations with
Ising-type, non-Abelian statistics.
We then generalize the formalism of extended Young
tableaux~\cite{greiter-07prl237202} to the case $S=1$ and find that
after every second spinon, there are two possible choices for the
quantization of the momentum spacing---it can either be integer or
half integer.  This yields an equivalent number of choices as we would
obtain if we had one Majorana fermion state for each spinon, in
analogy to the Majorana fermion states located at the quasiparticles
of the Moore--Read state in the quantum Hall
effect~\cite{moore-91npb362,kopnin-91prb9667,read-00prb10267}.  The
choice we have after the 1st, 3rd, 5th etc.\ spinon is that the
spacings between the neighboring single spinon momenta can either be
that of Bose (or Fermi) statistics, or be that of half-Fermi
statistics.  We call the momentum spacings for this model Majorana
spacings.  In Section \ref{sec:naa}, we review the generalization of
the model to arbitrary spin $S$, generalize the formalism of extended
Young tableaux accordingly, and derive the rules for the momentum
spacings of non-Abelian SU(2) level $k=2S$ anyons.  These are not as
easily stated, but reasonably simple as far as the principles are
concerned.  The mapping of these rules back to the related family of
quantized Hall states, the Read-Rezayi states~\cite{read-99prb8084},
may be instructive in formulating explicit matrix representations of
SU(2) level $k$ anyons in two dimensions, a task which so far has been
accomplished only for Ising anyons~\cite{ivanov01prl268}.  Finally, we
establish a one-to-one correspondence between the rules for the
momentum spacings we have derived from the formalism of extended Young
tableaux, and the fusion rules of $j=\frac{1}{2}$ spinons within the
SU(2) level $k=2S$ algebra.  The physical information
regarding the momentum spacings can hence be transferred to the spinon
bases~\cite{bouwknegt-94plb448,bouwknegt-95plb304,bouwknegt-96npb345,bouwknegt-99npb501}
of the conformal field theory, the Wess--Zumino--Witten
model~\cite{wess-71plb95,witten84cmp455,affleck86npb409,affleck-87prb5291}.
 
\section{Fractional momentum spacings and Abelian anyons in 1D}
\label{sec:hs}

\subsection{The Haldane--Shastry model}
\label{sec:hsmod}
The Haldane--Shastry
model~\cite{haldane88prl635,shastry88prl639,haldane91prl1529,shastry92prl164,haldane-92prl2021,kawakami92prb1005,talstra95,bernevig-01prl3392,greiter-06prl059701,%
greiter-07prl237202}
is one of the most important paradigms for a generic spin \half liquid
on a chain.  Consider a spin \half chain with periodic boundary
conditions and an even number of sites $N$ on a unit circle embedded
in the complex plane:
\begin{center}
\begin{picture}(310,45)(-30,-25)
\put(0,0){\circle{100}}
\put( 20.0,   .0){\circle*{3}}
\put( 17.3, 10.0){\circle*{3}}
\put( 10.0, 17.3){\circle*{3}}
\put(   .0, 20.0){\circle*{3}}
\put(-10.0, 17.3){\circle*{3}}
\put(-17.3, 10.0){\circle*{3}}
\put(-20.0,   .0){\circle*{3}}
\put(-17.3,-10.0){\circle*{3}}
\put(-10.0,-17.3){\circle*{3}}
\put(   .0,-20.0){\circle*{3}}
\put( 10.0,-17.3){\circle*{3}}
\put( 17.3,-10.0){\circle*{3}}
\qbezier[20]( 20.0,   .0)(  5.0,8.65)(-10.0, 17.3)
\put(50,8){\makebox(0,0)[l]{$N$\ sites with spin \half on unit circle:}}
\put(50,-8){\makebox(0,0)[l]
{$\displaystyle \eta_\alpha=e^{\i \frac{2\pi}{N}\alpha }$
\ \ with\ $\alpha = 1,\ldots ,N$}}
\end{picture}
\end{center}
The ${1}/{r^2}$-Hamiltonian
\begin{align}
  \label{eq:hsham}
  {H}^{\s\text{HS}} = \left(\frac{2\pi}{N}\right)^2
  \sum^N_{\alpha <\beta}\,
  \frac{{\boldsymbol{S}}_\alpha {\boldsymbol{S}}_\beta 
  }{\left|\eta_\alpha-\eta_\beta \right|^2}\,,
\end{align}
where $\left|\eta_\alpha-\eta_\beta \right|$ is the chord distance between
the sites $\alpha$ and $\beta$, has the exact ground state 
\begin{align}
  \label{eq:hsket}
  \ket{\psi^{\s\text{HS}}_{0}}=\hspace{-5pt} 
  \sum_{\{z_1,\ldots ,z_M\}}\hspace{-5pt}\psi^{\s\text{HS}}_{0} 
  (z_1,\ldots ,z_M)\,{S}^+_{z_1}\cdot\ldots\cdot {S}^+_{z_M} 
  \big|\underbrace{\dw\dw\ldots
    \dw}_{\hspace{-5pt} \text{all\ } N \text{\ spins\ } \dw \hspace{-5pt} }
  \big\rangle,\nonumber\\[3pt]  \\[-19pt]\nonumber
\end{align}
where the sum extends over all possible ways to distribute the 
$M=\frac{N}{2}$ $\up$-spin coordinates $z_i$ on the unit circle and
\begin{align}
  \label{eq:hspsi0}
  \psi^{\s\text{HS}}_{0}(z_1,z_2,\ldots ,z_M) =
  \prod_{i<i}^M\,(z_i-z_j)^2\,\prod_{i=1}^M\,z_i\,. 
\end{align}
The ground state is real, a spin singlet, has momentum
\begin{align}
  \label{eq:hsgp0}
  p_0=-\frac{\pi}{2} N,
\end{align}
where we have adopted a convention according to which the
``vacuum'' state $\ket{\dw\dw\ldots\dw}$ has momentum $p=0$ (and the
empty state $\ket{0}$ has $p=\pi (N-1)$), and energy
\begin{align}
  \label{eq:hse0}
  E_0=-\frac{\pi^2}{24}\left(N+\frac{5}{N}\right).
\end{align}
The ground state \eqref{eq:hsket} with \eqref{eq:hspsi0} was 
known long before the model, as it can be obtained by Gutzwiller projection
from Slater determinant states describing filled
bands~\cite{gutzwiller63prl159,kaplan-82prl889,metzner-87prl121}.
The Hamiltonian \eqref{eq:hsham} possesses a Yangian
symmetry~\cite{haldane-92prl2021,ha-93prb12459}, is fully
integrable~\cite{talstra-95jpa2369}, and also amenable to exact
solution via the asymtotic Bethe
Ansatz~\cite{haldane91prl1529,kawakami92prb1005,ha-93prb12459,ha-94prl2887}.

We will not verify the model explicitly here, but rather focus
on the fractional momentum spacings of the spinon excitations,
which reflect their Abelian anyon statistics.  

\subsection{Spinon excitations and fractional statistics}
\label{sec:hsspinons}

The elementary excitations for the Haldane--Shastry model are free
spinon excitations, which carry spin \half and no charge.  They
constitute an instance of fractional quantization, which is both
conceptually and mathematically similar to the fractional quantization
of charge in the fractional quantum Hall
effect~\cite{laughlin83prl1395}.  Their fractional quantum number is
the spin, which takes the value \half in a Hilbert space
(\ref{eq:hsket}) made out of spin flips ${S}^+$, which carry spin 1.

\vspace{\baselineskip}
\emph{One-spinon states.}---%
To write the wave function for a $\dw$-spin spinon localized at site 
$\eta_\alpha$, consider a chain with an odd number of sites $N$ and
let $M=\frac{N-1}{2}$ be the number of $\up$ or $\dw$ spins condensed
in the uniform liquid.  The spinon wave function is then given by
\begin{align}
  \label{eq:hsp1sp}
  \psi_{\alpha\dw}
  (z_1,z_2,\ldots ,z_M) =\prod_{i=1}^M\,(\eta_\alpha -z_i)\,
  \psi^{\s\text{HS}}_{0}(z_1,z_2,\ldots ,z_M),
\end{align}
which we understand substituted into (\ref{eq:hsket}).  
It is easy to verify 
${S}^\z _{\text{tot}} \psi_{\alpha\dw} = 
-\frac{1}{2} \psi_{\alpha\dw}$ and 
${S}^-_{\text{tot}} \psi_{\alpha\dw} = 0$,
which shows that the spinon transforms as a spinor under rotations.  

The localized spinon (\ref{eq:hsp1sp}) is not an eigenstate of the
Hamiltonian (\ref{eq:hsham}).  To obtain exact
eigenstates, we construct momentum eigenstates according to
\begin{align}
  \label{eq:hspsim}
  \psi_{m\dw}(z_1,z_2,\ldots ,z_M) =
  \sum_{\alpha=1}^{N} (\bar\eta_\alpha)^m\,
  \psi_{\alpha\dw}(z_1,z_2,\ldots ,z_M), 
\end{align}
where the integer $m$ corresponds to a momentum quantum number.
Since $\psi_{\alpha\dw} (z_1,z_2,\ldots ,z_M)$ contains
only powers $\eta_\alpha^0, \eta_\alpha^1,\ldots , \eta_\alpha^M$ and
\begin{align}
  \label{eq:hsdelta}
  \sum_{\alpha=1}^{N} \overline{\eta}_\alpha^m \eta_\alpha^n = \delta_{mn}
  \quad \text{mod}\ N,
\end{align}
$\psi_{m\dw} (z_1,z_2,\ldots ,z_M)$ will vanish unless $m=0,1,\ldots
,M$.  There are only roughly half as many-spinon orbitals as there are
sites.  Spinons on neighboring sites hence cannot be orthogonal.
Acting with \eqref{eq:hsham} on \eqref{eq:hspsim}, we
obtain~\cite{haldane91prl1529,bernevig-01prl3392,bernevig-01prb024425,Greiter11}
\begin{align}
  H^{\s\text{HS}}\ket{\psi_{m\dw}}
  =
  \left[ -\frac{\pi^2}{24}\left( N-\frac{1}{N}\right)+
    \frac{2\pi^2}{N^2} m(M-m) \right] \!\ket{\psi_{m\dw}}.
\label{eq:hs1spinonenergy}
\end{align}

To make a correspondence between $m$ and the spinon momentum $p_m$, we
translate (\ref{eq:hspsim}) counterclockwise by one lattice spacing
(which we set to unity for present purposes) around the unit circle,
\begin{align}
  \label{eq:hst}
  \boldsymbol{T}\, \ket{\psi_{m\dw}} 
  = e^{\text{i}(p_0+p_m)}\ket{\psi_{m\dw}}.
\end{align}
With $p_0=-\frac{\pi}{2}N$, we find
\begin{align}
  \label{eq:hsq}
  p_m=\pi-\frac{2\pi}{N}\!\left(m+\frac{1}{4}\right).
\end{align}

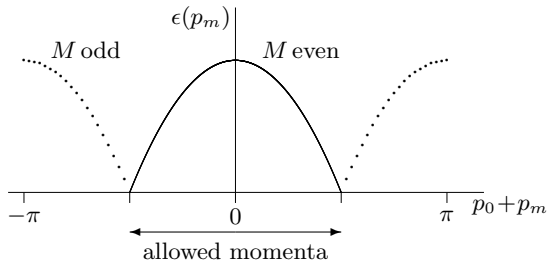
\begin{figure}[tb]
  \begin{center}
    \begin{picture}(200,110)(-100,-30)
      \qbezier[2000](-40,0)(0,100)(40,0)
      \put(4,0){\makebox(0,0){\rule{180.pt}{0.3pt}}}
      \put(0,0){\makebox(0,0)[b]{\rule{0.3pt}{70.pt}}}
      \put(-80,0){\makebox(0,0)[t]{\rule{0.3pt}{4pt}}}
      \put(-80,-9){\makebox(0,0){\small $-\pi$}}
      \put(-40,0){\makebox(0,0)[t]{\rule{0.3pt}{4pt}}}
      \put(0,0){\makebox(0,0)[t]{\rule{0.3pt}{4pt}}}
      \put(0,-9){\makebox(0,0){\small 0}}
      \put(40,0){\makebox(0,0)[t]{\rule{0.3pt}{4pt}}}
      \put(80,0){\makebox(0,0)[t]{\rule{0.3pt}{4pt}}}
      \put(80,-9){\makebox(0,0){\small $\pi$}}
      \put(104,-5){\makebox(0,0){\small $p_0\!+\!p_m$}}
      \put(-13,66){\makebox(0,0){\small $\epsilon(p_m)$}}
      \put(0,-15){{\vector(-1,0){40}}}
      \put(0,-15){{\vector(1,0){40}}}
      \put(0,-22){\makebox(0,0){\small allowed momenta}}
      \put(10,54){\makebox(0,0)[l]{\small $M\!$ even}}
      \put(-70,54){\makebox(0,0)[l]{\small $M\!$ odd}}
      \put(-80.0, 50.0){\circle*{1}}
      \put(-78.0, 49.9){\circle*{1}}
      \put(-76.0, 49.5){\circle*{1}}
      \put(-74.0, 48.9){\circle*{1}}
      \put(-72.0, 48.0){\circle*{1}}
      \put(-70.0, 46.9){\circle*{1}}
      \put(-68.0, 45.5){\circle*{1}}
      \put(-66.0, 43.9){\circle*{1}}
      \put(-64.0, 42.0){\circle*{1}}
      \put(-62.0, 39.9){\circle*{1}}
      \put(-60.0, 37.5){\circle*{1}}
      \put(-58.0, 34.9){\circle*{1}}
      \put(-56.0, 32.0){\circle*{1}}
      \put(-54.0, 28.9){\circle*{1}}
      \put(-52.0, 25.5){\circle*{1}}
      \put(-50.0, 21.9){\circle*{1}}
      \put(-48.0, 18.0){\circle*{1}}
      \put(-46.0, 13.9){\circle*{1}}
      \put(-44.0,  9.5){\circle*{1}}
      \put(-42.0,  4.9){\circle*{1}}
      \put( 80.0, 50.0){\circle*{1}}
      \put( 78.0, 49.9){\circle*{1}}
      \put( 76.0, 49.5){\circle*{1}}
      \put( 74.0, 48.9){\circle*{1}}
      \put( 72.0, 48.0){\circle*{1}}
      \put( 70.0, 46.9){\circle*{1}}
      \put( 68.0, 45.5){\circle*{1}}
      \put( 66.0, 43.9){\circle*{1}}
      \put( 64.0, 42.0){\circle*{1}}
      \put( 62.0, 39.9){\circle*{1}}
      \put( 60.0, 37.5){\circle*{1}}
      \put( 58.0, 34.9){\circle*{1}}
      \put( 56.0, 32.0){\circle*{1}}
      \put( 54.0, 28.9){\circle*{1}}
      \put( 52.0, 25.5){\circle*{1}}
      \put( 50.0, 21.9){\circle*{1}}
      \put( 48.0, 18.0){\circle*{1}}
      \put( 46.0, 13.9){\circle*{1}}
      \put( 44.0,  9.5){\circle*{1}}
      \put( 42.0,  4.9){\circle*{1}}
    \end{picture}
    \caption{Dispersion of a single spinon in a Haldane--Shastry chain.}
    \label{fig:hs1spd}
  \end{center}
\end{figure}

The energy \eqref{eq:hs1spinonenergy} can be written as
$E=E_0+\epsilon(p_m)$, with the spinon dispersion given by
\begin{align}
  \label{eq:hsep}
  \epsilon(p)=\frac{1}{2}p\left(\pi-p\right)+\frac{\pi^2}{8N^2},
\end{align}
as depicted in Figure \ref{fig:hs1spd}.  The interval of allowed spinon
momenta spans only half of the Brillouin zone, and alternates with $M$
even vs.\ $M$ odd.


\vspace{\baselineskip}
\emph{Two-spinon states.}---%
To write the wave function for two $\dw$-spin spinons localized at sites 
$\eta_\alpha$ and $\eta_\beta$, consider a chain with $N$ even and 
$M=\frac{N-2}{2}$.  The two-spinon state is then given by
\begin{align}
  \label{eq:hsp2sp}
  \psi_{\alpha\beta}(z_1,\ldots ,z_M) =
  \prod_{i=1}^M\,(\eta_\alpha -z_i)(\eta_\beta -z_i)\,
  \psi^{\s\text{HS}}_{0}(z_1,\ldots ,z_M).
\end{align}
A momentum basis for the two-spinon states is given by
\begin{align}
  \label{eq:hspsimn}
  \psi_{mn}(z_1,\ldots ,z_M) =
  \sum_{\alpha,\beta=1}^{N} (\bar\eta_\alpha)^m\,(\bar\eta_\beta)^m\,
  \psi_{\alpha\beta}(z_1,\ldots ,z_M), 
\end{align}
where $M\ge m\ge n\ge 0$.  For $m$ or $n$ outside this range,
$\psi_{mn}$ vanishes identically, reflecting the overcompleteness of
the position space basis.
Acting with \eqref{eq:hsham} on \eqref{eq:hspsim}, we
obtain~\cite{haldane91prl1529,bernevig-01prl3392,bernevig-01prb024425,Greiter11}
\begin{align}
  \label{eq:hsscattspinon}
  H^{\s\text{HS}}\ket{\psi_{mn}}=
  E_{mn}\ket{\psi_{mn}}+\sum_{l=1}^{l_{\text{max}}}V_l^{mn}\ket{\psi_{m+l,n-l}}
\end{align}
with
%
\begin{align}
  \label{eq:hs2spE}
  E_{mn}=&\textstyle
  -\frac{\pi^2}{24}\left(N-\frac{19}{N}+\frac{24}{N^2}\right)
  \nonumber\\[2pt]
  &\textstyle
  +\frac{2\pi^2}{N^2}\Bigl[ m\left(\frac{N}{2} - 1 - m\right) 
  + n\left(\frac{N}{2} - 1 - n\right)- \frac{m-n}{2}\Bigr],
  \\
  V_l^{mn}=&\textstyle -\frac{2\pi^2}{N^2}(m-n+2l),
\end{align}
and $l_{\text{max}}=\min(M-m,n)$.  Since the ``scattering'' of the
non-ortho\-gonal basis states $\ket{\psi_{mn}}$ in
\eqref{eq:hsscattspinon} only occurs in one direction, increasing
$m-n$ while keeping $m+n$ fixed, the eigenstates of $H^{\s\text{HS}}$
have energy eigenvalues $E_{mn}$, and are of the form
\begin{align}
  \label{eq:hsphinm}
  \ket{\phi_{mn}}
  =\sum_{l=0}^{l_M} a_l^{mn}\ket{\psi_{m+l,n-l}}.
\end{align}
A recursion relation for the coefficients $a_l^{mn}$ is readily 
obtained from \eqref{eq:hsscattspinon}.

If we identify the single-spinon momenta for $m\ge n$ according
to
\begin{align}
  \label{eq:hsqmqnspinonpm}
  p_m&=\pi-\frac{2\pi}{N}\left(m+\frac{1}{2}+s\right),
  \\[4pt]\label{eq:hsqmqnspinonpn}
  p_n&=\pi-\frac{2\pi}{N}\left(n+\frac{1}{2}-s\right),
\end{align}
with a {statistical shift}
$s=\frac{1}{4}$~\cite{greiter-05prb224424,greiter-06prl059701}, we can write
the energy
\begin{align}
  \label{eq:hsetotspinon}
  E_{mn}=E_0+\epsilon(p_m)+\epsilon(p_n),
\end{align}
where $E_0$ is the ground state energy \eqref{eq:hse0} and
$\epsilon(p)$ the spinon dispersion \eqref{eq:hsep}.

\vspace{\baselineskip}
\emph{Fractional statistics.}---%
The mutual half-Fermi statistics of the spinons manifests itself in
the fractional shift $s$ in the
single-spinon momenta \eqref{eq:hsqmqnspinonpm} and \eqref{eq:hsqmqnspinonpn}, as we will elaborate 
now~\cite{greiter09prb064409}.  
The Ansatz \eqref{eq:hspsimn} unambiguously implies that the sum of
the two spinon momenta is given by
$q_m+q_n=2\pi-\frac{2\pi}{N}(m+n+1)$, and hence
\eqref{eq:hsqmqnspinonpm} and \eqref{eq:hsqmqnspinonpn}.  The shift
$s$ is determined by demanding that the excitation energy
\eqref{eq:hsetotspinon} of the two-spinon state is a sum of
single-spinon energies, which in turn is required for the explicit
solution here to be consistent with the models solution via the
asymptotic Bethe
ansatz~\cite{ha-93prb12459,essler95prb13357,greiter-05prb224424}.

\begin{figure}[tb]
\begin{center}
  \begin{picture}(280,65)(-23,-20) \put(0,0){\circle{100}} \put(
    17.3,10.0){\circle*{4}} \put(-10.0, 17.3){\circle*{4}}
    \put(26.8,14){\makebox(0,0)[l]{\vector(-2,3){7}}}
    \put(-18,24){\makebox(0,0)[l]{\vector(3,2){10}}}
    \put(26,18){\makebox(0,0)[l]{\small $1$}}
    \put(-19,30){\makebox(0,0)[l]{\small $2$}}
    \put(45,32){\makebox(0,0)[l] {\small relative motion of
        one-dimensional anyons}}
    \put(45,22){\makebox(0,0)[l] {\small is unidirectional 
        (e.g.\ 2 moves clockwise}}
    \put(45,12){\makebox(0,0)[l]{\small relative to 1)}}
    \put(45,-3){\makebox(0,0)[l]{\small when
        anyons cross:\quad $|\psi\!> \rightarrow e^{\text{i}\theta}
        |\psi\!>$}} 
    \put(45,-20){\makebox(0,0)[l]{\small mom.\
        spacing:\quad $\displaystyle p_1\!-\!p_2 = \Delta p\rightarrow
        \Delta p - \frac{2\hbar\theta}{L}$}}
  \end{picture}
  \end{center}
  \caption{Fractional statistics in one dimension. The crossings of
    the anyons are unidirectional, and the many particle wave function
    acquires a statistical phase $\theta$ whenever they cross.}
  \label{fig:statistics}
\end{figure}
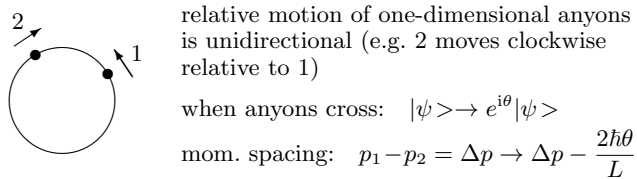

The shift decreases the momentum $p_m$ of spinon 1 and increases
momentum $p_n$ of spinon 2.  This may surprise at first as the basis
states \eqref{eq:hspsimn} are constructed symmetrically with regard to
interchanges of $m$ and $n$.  To understand the asymmetry, note that
$M\ge m\ge n\ge 0$ implies $0<p_m<p_n<\pi$.  The dispersion
 \eqref{eq:hsep} implies that the group velocity of the spinons is
given by
\begin{align}
  v_\text{g}(p)=\partial_p\epsilon(p)=\frac{\pi}{2}-p,
\end{align}
%
which in turn implies that $v_\text{g}(p_m)>v_\text{g}(p_n)$.  
This means that the \emph{relative motion} of spinon 1 (with $q_m$) with
respect to spinon 2 (with $q_n$) is \emph{always counterclockwise} on
the unit circle 
(see Figure \ref{fig:statistics}).  
The shifts in the individual spinon momenta can
hence be explained by assuming that the two-spinon state
acquires a statistical phase $\theta=2\pi s$ whenever the spinons pass
through each other.  This phase implies that $q_m$ is shifted by
$-\frac{2\pi}{N}s$ since we have to translate spinon 1
counterclockwise through spinon 2 and hence counterclockwise around
the unit circle when obtaining the allowed values for $q_m$ from the
PBCs.  Similarly, $q_n$ is shifted by $+\frac{2\pi}{N}s$ since we have
to translate spinon 2 clockwise through spinon 1 and hence clockwise
around the unit circle when obtaining the quantization of $q_n$.

That the crossing of the spinons occurs only in one direction is a
necessary requirement for fractional statistics to exist in one
dimension.  If the spinons could cross in both directions, the fact
that paths interchanging them twice (\ie once in each direction) are
topologically equivalent to paths not interchanging them at all would
imply $2\theta=0\ \text{mod}\ 2\pi$ for the statistical phase, \ie
only allow for the familiar choices of bosons or fermions.  With the
scattering occurring in only one direction, arbitrary values for
$\theta$ are possible.  Note that the one-dimensional anyons break
neither time-reversal symmetry (T) nor parity (P).

\vspace{.5\baselineskip} The fractional statistics of the spinons
manifests itself further in the fractional exclusion (or generalized
Pauli) principle introduced by one of us~\cite{haldane91prl937}.  If
we consider a state with $L$ spinons, we can easily see from
\eqref{eq:hspsim}, \eqref{eq:hsdelta}, and \eqref{eq:hspsimn} that the
number of orbitals available for further spinons we may wish to create
is $M+1$, where $M=\frac{N-L}{2}$ is the number of $\up$ or $\dw$
spins in the remaining uniform liquid.  (In this representation, the
spinon wave functions are symmetric; two or more spinons can have the
same value for $m$.)  In other words, the creation of {\em two}
spinons reduces the number of available single spinon states by {\em
  one}.  They hence obey half-Fermi statistics in the sense of the
generalized Pauli exclusion principle.  (For fermions, the creation of
two particles would decrease the number of available single particle
by two, while this number would not change for bosons.)


\subsection{Young tableaux and many-spinon states}
\label{sec:hsyt} 

The easiest way to obtain the spectrum of the model is through the
one-to-one correspondence between the Young tableaux classifying the
total spin representations of $N$ spins and the exact eigenstates of
the the Haldane--Shastry model for a chain with $N$ sites, which are
classified by the total spins and the fractionally spaced
single-particle momenta of the spinons~\cite{greiter-07prl237202}.

This correspondence yields the allowed sequences of single-spinon
momenta $p_1,\ldots,p_L$ as well as the allowed representations for
the total spin of the states such that the eigenstates of the Haldane--Shastry model have momenta and energies
\begin{align}
  p=p_0+\sum_{i=1}^L p_i,\quad E=E_0+\sum_{i=1}^L \epsilon(p_i),
  \label{eq:hsLspinonenergy}
\end{align} 
where $p_0$ and $E_0$ denote the ground state momentum and energy,
respectively, and $\epsilon(p)$ is the single-spinon dispersion.  The
correspondence hence does not only provide the quantum numbers of all
the states in the spectrum, but also shows that it is sensible to view
the individual spinons as particles, rather than just as solitons or
collective excitations in many body condensates.  We now proceed by
stating these rules without further motivating or even deriving them.

\begin{figure}[tb]
  \begin{center}
    \setlength{\unitlength}{\ytlength}
    \begin{picture}(32,8)(1.5,-1) \linethickness{0.5pt}
      \put(3,6){\line(1,0){1}} \put(3,5){\line(1,0){1}}
      \put(3,5){\line(0,1){1}} \put(4,5){\line(0,1){1}}
      \put(3,5){\makebox(1,1){1}} \put(4.6,5.3){$\otimes$}
      \put(6,6){\line(1,0){1}} \put(6,5){\line(1,0){1}}
      \put(6,5){\line(0,1){1}} \put(7,5){\line(0,1){1}}
      \put(6,5){\makebox(1,1){2}}
      \put(3,4.4){$\underbrace{\phantom{iiiiiiiiii}}$}
      \put(3,2.9){\line(1,0){1}} \put(3,1.9){\line(1,0){1}}
      \put(3,0.9){\line(1,0){1}} \put(3,0.9){\line(0,1){2}}
      \put(4,0.9){\line(0,1){2}} \put(3,1.9){\makebox(1,1){1}}
      \put(3,0.9){\makebox(1,1){2}} \put(2,-0.5){$S\!=\!0$}
      \put(4.6,1.5){$\oplus$} \put(6,2.4){\line(1,0){2}}
      \put(6,1.4){\line(1,0){2}} \put(6,1.4){\line(0,1){1}}
      \put(7,1.4){\line(0,1){1}} \put(8,1.4){\line(0,1){1}}
      \put(6,1.4){\makebox(1,1){1}} \put(7,1.4){\makebox(1,1){2}}
      \put(5.7,-0.2){$S\!=\!1$} \put(7.6,5.3){$\otimes$}
      \put(9,6){\line(1,0){1}} \put(9,5){\line(1,0){1}}
      \put(9,5){\line(0,1){1}} \put(10,5){\line(0,1){1}}
      \put(9,5){\makebox(1,1){3}} \put(11.6,5.3){$=$}
      \put(14,6){\line(1,0){1}} \put(14,5){\line(1,0){1}}
      \put(14,4){\line(1,0){1}} \put(14,3){\line(1,0){1}}
      \put(14,3){\line(0,1){3}} \put(15,3){\line(0,1){3}}
      \put(14,5){\makebox(1,1){1}} \put(14,4){\makebox(1,1){2}}
      \put(14,3){\makebox(1,1){3}} \put(16.1,5.3){$\oplus$}
      \put(18,6){\line(1,0){2}} \put(18,5){\line(1,0){2}}
      \put(18,4){\line(1,0){1}} \put(18,4){\line(0,1){2}}
      \put(19,4){\line(0,1){2}} \put(20,5){\line(0,1){1}}
      \put(18,5){\makebox(1,1){1}} \put(19,5){\makebox(1,1){2}}
      \put(18,4){\makebox(1,1){3}} \put(17.5,2.2){$S\!=\!\frac{1}{2}$}
      \put(21.1,5.3){$\oplus$} \put(23,6){\line(1,0){2}}
      \put(23,5){\line(1,0){2}} \put(23,4){\line(1,0){1}}
      \put(23,4){\line(0,1){2}} \put(24,4){\line(0,1){2}}
      \put(25,5){\line(0,1){1}} \put(23,5){\makebox(1,1){1}}
      \put(23,4){\makebox(1,1){2}} \put(24,5){\makebox(1,1){3}}
      \put(22.5,2.2){$S\!=\!\frac{1}{2}$} \put(26.1,5.3){$\oplus$}
      \put(28,6){\line(1,0){3}} \put(28,5){\line(1,0){3}}
      \put(28,5){\line(0,1){1}} \put(29,5){\line(0,1){1}}
      \put(30,5){\line(0,1){1}} \put(31,5){\line(0,1){1}}
      \put(28,5){\makebox(1,1){1}} \put(29,5){\makebox(1,1){2}}
      \put(30,5){\makebox(1,1){3}} \put(28,2.2){$S\!=\!\frac{3}{2}$}
      \put(14.5,4.5){\makebox(0,0){\line(1,2){1.8}}}
      \put(14.5,4.5){\makebox(0,0){\line(1,-2){1.8}}}
    \end{picture}
    \caption{Total spin representations of three $S=\frac{1}{2}$ spins
      with Young tableaux. For SU($n$) with $n>2$, the tableaux with
      three boxes on top of each other would exist as well.}
    \label{fig:hsyoungdiagram}
  \end{center}
\end{figure}

To begin with, the Hilbert space of a system of $N$ identical SU($n$)
spins can be decomposed into representations of the total spin, which
commutes with \eqref{eq:hsham} and hence can be used to classify the
eigenstates.  
These representations are compatible with the representations of the
symmetric group S$_N$ of $N$ elements, which may be expressed in terms
of Young tableaux~\cite{Hamermesh62,InuiTanabeOnodera96}.  The general
rule for obtaining Young tableaux is illustrated for three
$S=\frac{1}{2}$ spins in Fig.~\ref{fig:hsyoungdiagram}.
For each of the $N$ spins, draw a box and 
number the boxes consecutively from left to right.  The representations of
SU($n$) are constructed by putting the boxes together such that the
numbers assigned to them increase in each row from left to right and
in each column from top to bottom.  Each tableau indicates
symmetrization over all boxes in the same row, and antisymmetrization
over all boxes in the same column.  This implies that we cannot have
more than $n$ boxes on top of each other for SU($n$) spins.  For
SU(2), each tableau corresponds to a spin
$S=\frac{1}{2}(\lambda_1-\lambda_2)$ representation, with $\lambda_i$
the number of boxes in the $i\,\text{th}$ row, and stands for a
multiplet $S^\z =-S,\ldots,S$.

\begin{figure*}[tb]
  \begin{center}
    \setlength{\unitlength}{\ytlength}
    \begin{picture}(32,6)(-.6,0) \linethickness{0.5pt}
      \put(5.25,4){\makebox(1,1){$S_{\text{tot}}$}}
      \put(21.5,4){\makebox(1,1){$L$}} \put(24,4.1){$a_1,\dots,a_L$}
      \put(0,3){\line(1,0){2}} \put(0,2){\line(1,0){2}}
      \put(0,1){\line(1,0){2}} \put(0,1){\line(0,1){2}}
      \put(1,1){\line(0,1){2}} \put(2,1){\line(0,1){2}}
      \put(0,2){\makebox(1,1){1}} \put(0,1){\makebox(1,1){2}}
      \put(1,2){\makebox(1,1){3}} \put(1,1){\makebox(1,1){4}}
      \put(5.25,1.5){\makebox(1,1){0}} \put(7,1.8){$\rightarrow$}
      \put(9,3){\line(1,0){2}} \put(9,2){\line(1,0){2}}
      \put(9,1){\line(1,0){2}} \put(9,1){\line(0,1){2}}
      \put(10,1){\line(0,1){2}} \put(11,1){\line(0,1){2}}
      \put(9,2){\makebox(1,1){1}} \put(9,1){\makebox(1,1){2}}
      \put(10,2){\makebox(1,1){3}} \put(10,1){\makebox(1,1){4}}
      \put(14,1.8){$\rightarrow$} \put(16,3){\line(1,0){2}}
      \put(16,2){\line(1,0){2}} \put(16,1){\line(1,0){2}}
      \put(16,1){\line(0,1){2}} \put(17,1){\line(0,1){2}}
      \put(18,1){\line(0,1){2}} \put(16,2){\makebox(1,1){1}}
      \put(16,1){\makebox(1,1){2}} \put(17,2){\makebox(1,1){3}}
      \put(17,1){\makebox(1,1){4}} \put(21.5,1.5){\makebox(1,1){0}}
      \put(24,2){\line(1,0){5}}
      \multiput(25,1.85)(1,0){4}{\rule{0.5pt}{2pt}}
      \put(31,3.7){\makebox(1,1){$p_{\text{tot}}$}}
      \put(31,1.5){\makebox(1,1){$0$}}
    \end{picture}

    \begin{picture}(32,3.5)(-.6,0) \linethickness{0.5pt}
      \put(0,3){\line(1,0){2}} \put(0,2){\line(1,0){2}}
      \put(0,1){\line(1,0){2}} \put(0,1){\line(0,1){2}}
      \put(1,1){\line(0,1){2}} \put(2,1){\line(0,1){2}}
      \put(0,2){\makebox(1,1){1}} \put(1,2){\makebox(1,1){2}}
      \put(0,1){\makebox(1,1){3}} \put(1,1){\makebox(1,1){4}}
      \put(5.25,1.5){\makebox(1,1){0}} \put(7,1.8){$\rightarrow$}
      \put(9,3){\line(1,0){2}} \put(9,2){\line(1,0){3}}
      \put(10,1){\line(1,0){2}} \put(9,2){\line(0,1){1}}
      \put(10,1){\line(0,1){2}} \put(11,1){\line(0,1){2}}
      \put(12,1){\line(0,1){1}} \put(9,2){\makebox(1,1){1}}
      \put(10,2){\makebox(1,1){2}} \put(10,1){\makebox(1,1){3}}
      \put(11,1){\makebox(1,1){4}} \put(14,1.8){$\rightarrow$}
      \put(16,3){\line(1,0){2}} \put(16,2){\line(1,0){3}}
      \put(17,1){\line(1,0){2}} \put(16,2){\line(0,1){1}}
      \put(17,1){\line(0,1){2}} \put(18,1){\line(0,1){2}}
      \put(19,1){\line(0,1){1}} \put(16,2){\makebox(1,1){1}}
      \put(17,2){\makebox(1,1){2}} \put(17,1){\makebox(1,1){3}}
      \put(18,1){\makebox(1,1){4}} \put(16.5,1.5){\circle*{0.4}}
      \put(18.5,2.5){\circle*{0.4}} \put(21.5,1.5){\makebox(1,1){2}}
      \put(24,2){\line(1,0){5}}
      \multiput(25,1.85)(1,0){4}{\rule{0.5pt}{2pt}}
      \multiput(25,2)(3,0){2}{\circle*{0.5}}
      \put(24.5,0.5){\makebox(1,1){1}}
      \put(27.5,0.5){\makebox(1,1){4}}
      \put(31,1.5){\makebox(1,1){$\pi$}}
    \end{picture}

    \begin{picture}(32,3.5)(-.6,0) \linethickness{0.5pt}
      \put(0,3){\line(1,0){3}} \put(0,2){\line(1,0){3}}
      \put(0,1){\line(1,0){1}} \put(0,1){\line(0,1){2}}
      \put(1,1){\line(0,1){2}} \put(2,2){\line(0,1){1}}
      \put(3,2){\line(0,1){1}} \put(0,2){\makebox(1,1){1}}
      \put(0,1){\makebox(1,1){2}} \put(1,2){\makebox(1,1){3}}
      \put(2,2){\makebox(1,1){4}} \put(5.25,1.5){\makebox(1,1){1}}
      \put(7,1.8){$\rightarrow$} \put(9,3){\line(1,0){3}}
      \put(9,2){\line(1,0){3}} \put(9,1){\line(1,0){1}}
      \put(9,1){\line(0,1){2}} \put(10,1){\line(0,1){2}}
      \put(11,2){\line(0,1){1}} \put(12,2){\line(0,1){1}}
      \put(9,2){\makebox(1,1){1}} \put(9,1){\makebox(1,1){2}}
      \put(10,2){\makebox(1,1){3}} \put(11,2){\makebox(1,1){4}}
      \put(14,1.8){$\rightarrow$} \put(16,3){\line(1,0){3}}
      \put(16,2){\line(1,0){3}} \put(16,1){\line(1,0){1}}
      \put(16,1){\line(0,1){2}} \put(17,1){\line(0,1){2}}
      \put(18,2){\line(0,1){1}} \put(19,2){\line(0,1){1}}
      \put(16,2){\makebox(1,1){1}} \put(16,1){\makebox(1,1){2}}
      \put(17,2){\makebox(1,1){3}} \put(18,2){\makebox(1,1){4}}
      \put(17.5,1.5){\circle*{0.4}} \put(18.5,1.5){\circle*{0.4}}
      \put(21.5,1.5){\makebox(1,1){2}} \put(24,2){\line(1,0){5}}
      \multiput(25,103.85)(1,0){4}{\rule{0.5pt}{2pt}}
      \multiput(27,2)(1,0){2}{\circle*{0.5}}
      \put(26.5,0.5){\makebox(1,1){3}}
      \put(27.5,0.5){\makebox(1,1){4}}
      \put(31,1.5){\makebox(1,1){$\frac{3\pi}{2}$}}
    \end{picture}

    \begin{picture}(32,3.5)(-.6,0) \linethickness{0.5pt}
      \put(0,3){\line(1,0){3}} \put(0,2){\line(1,0){3}}
      \put(0,1){\line(1,0){1}} \put(0,1){\line(0,1){2}}
      \put(1,1){\line(0,1){2}} \put(2,2){\line(0,1){1}}
      \put(3,2){\line(0,1){1}} \put(0,2){\makebox(1,1){1}}
      \put(1,2){\makebox(1,1){2}} \put(0,1){\makebox(1,1){3}}
      \put(2,2){\makebox(1,1){4}} \put(5.25,1.5){\makebox(1,1){1}}
      \put(7,1.8){$\rightarrow$} \put(9,3){\line(1,0){3}}
      \put(9,2){\line(1,0){3}} \put(10,1){\line(1,0){1}}
      \put(9,2){\line(0,1){1}} \put(10,1){\line(0,1){2}}
      \put(11,1){\line(0,1){2}} \put(12,2){\line(0,1){1}}
      \put(9,2){\makebox(1,1){1}} \put(10,2){\makebox(1,1){2}}
      \put(10,1){\makebox(1,1){3}} \put(11,2){\makebox(1,1){4}}
      \put(14,1.8){$\rightarrow$} \put(16,3){\line(1,0){3}}
      \put(16,2){\line(1,0){3}} \put(17,1){\line(1,0){1}}
      \put(16,2){\line(0,1){1}} \put(17,1){\line(0,1){2}}
      \put(18,1){\line(0,1){2}} \put(19,2){\line(0,1){1}}
      \put(16,2){\makebox(1,1){1}} \put(17,2){\makebox(1,1){2}}
      \put(17,1){\makebox(1,1){3}} \put(18,2){\makebox(1,1){4}}
      \put(16.5,1.5){\circle*{0.4}} \put(18.5,1.5){\circle*{0.4}}
      \put(21.5,1.5){\makebox(1,1){2}} \put(24,2){\line(1,0){5}}
      \multiput(25,1.85)(1,0){4}{\rule{0.5pt}{2pt}}
      \multiput(25,2)(3,0){2}{\circle*{0.5}}
      \put(24.5,0.5){\makebox(1,1){1}}
      \put(27.5,0.5){\makebox(1,1){4}}
      \put(31,1.5){\makebox(1,1){$\pi$}}
    \end{picture}

    \begin{picture}(32,3.5)(-.6,0) \linethickness{0.5pt}
      \put(0,3){\line(1,0){3}} \put(0,2){\line(1,0){3}}
      \put(0,1){\line(1,0){1}} \put(0,1){\line(0,1){2}}
      \put(1,1){\line(0,1){2}} \put(2,2){\line(0,1){1}}
      \put(3,2){\line(0,1){1}} \put(0,2){\makebox(1,1){1}}
      \put(1,2){\makebox(1,1){2}} \put(2,2){\makebox(1,1){3}}
      \put(0,1){\makebox(1,1){4}} \put(5.25,1.5){\makebox(1,1){1}}
      \put(7,1.8){$\rightarrow$} \put(9,3){\line(1,0){3}}
      \put(9,2){\line(1,0){3}} \put(11,1){\line(1,0){1}}
      \put(9,2){\line(0,1){1}} \put(10,2){\line(0,1){1}}
      \put(11,1){\line(0,1){2}} \put(12,1){\line(0,1){2}}
      \put(9,2){\makebox(1,1){1}} \put(10,2){\makebox(1,1){2}}
      \put(11,2){\makebox(1,1){3}} \put(11,1){\makebox(1,1){4}}
      \put(14,1.8){$\rightarrow$} \put(16,3){\line(1,0){3}}
      \put(16,2){\line(1,0){3}} \put(18,1){\line(1,0){1}}
      \put(17,2){\line(0,1){1}} \put(16,2){\line(0,1){1}}
      \put(18,1){\line(0,1){2}} \put(19,1){\line(0,1){2}}
      \put(16,2){\makebox(1,1){1}} \put(17,2){\makebox(1,1){2}}
      \put(18,2){\makebox(1,1){3}} \put(18,1){\makebox(1,1){4}}
      \put(16.5,1.5){\circle*{0.4}} \put(17.5,1.5){\circle*{0.4}}
      \put(21.5,1.5){\makebox(1,1){2}} \put(24,2){\line(1,0){5}}
      \multiput(25,1.85)(1,0){4}{\rule{0.5pt}{2pt}}
      \multiput(25,2)(1,0){2}{\circle*{0.5}}
      \put(24.5,0.5){\makebox(1,1){1}}
      \put(25.5,0.5){\makebox(1,1){2}}
      \put(31,1.5){\makebox(1,1){$\frac{\pi}{2}$}}
    \end{picture}

    \begin{picture}(32,3.5)(-.6,0) \linethickness{0.5pt}
      \put(0,3){\line(1,0){4}} \put(0,2){\line(1,0){4}}
      \put(0,2){\line(0,1){1}} \put(1,2){\line(0,1){1}}
      \put(2,2){\line(0,1){1}} \put(3,2){\line(0,1){1}}
      \put(4,2){\line(0,1){1}} \put(0,2){\makebox(1,1){1}}
      \put(1,2){\makebox(1,1){2}} \put(2,2){\makebox(1,1){3}}
      \put(3,2){\makebox(1,1){4}} \put(5.25,1.5){\makebox(1,1){2}}
      \put(7,1.8){$\rightarrow$} \put(9,3){\line(1,0){4}}
      \put(9,2){\line(1,0){4}} \put(9,2){\line(0,1){1}}
      \put(10,2){\line(0,1){1}} \put(11,2){\line(0,1){1}}
      \put(12,2){\line(0,1){1}} \put(13,2){\line(0,1){1}}
      \put(9,2){\makebox(1,1){1}} \put(10,2){\makebox(1,1){2}}
      \put(11,2){\makebox(1,1){3}} \put(12,2){\makebox(1,1){4}}
      \put(14,1.8){$\rightarrow$} \put(16,3){\line(1,0){4}}
      \put(16,2){\line(1,0){4}} \put(16,2){\line(0,1){1}}
      \put(17,2){\line(0,1){1}} \put(18,2){\line(0,1){1}}
      \put(19,2){\line(0,1){1}} \put(20,2){\line(0,1){1}}
      \put(16,2){\makebox(1,1){1}} \put(17,2){\makebox(1,1){2}}
      \put(18,2){\makebox(1,1){3}} \put(19,2){\makebox(1,1){4}}
      \put(16.5,1.5){\circle*{0.4}} \put(17.5,1.5){\circle*{0.4}}
      \put(18.5,1.5){\circle*{0.4}} \put(19.5,1.5){\circle*{0.4}}
      \put(21.5,1.5){\makebox(1,1){4}} \put(24,2){\line(1,0){5}}
      \multiput(25,2)(1,0){4}{\circle*{0.5}}
      \put(24.5,0.5){\makebox(1,1){1}}
      \put(25.5,0.5){\makebox(1,1){2}}
      \put(26.5,0.5){\makebox(1,1){3}}
      \put(27.5,0.5){\makebox(1,1){4}}
      \put(31,1.5){\makebox(1,1){$0$}}
    \end{picture}
    \caption{Young tableau decomposition and the corresponding spinon
      states for an $S=\frac{1}{2}$ spin chain with $N=4$ sites.  The
      dots represent the spinons.  The spinon momentum
      numbers $a_i$ are given by the numbers in the boxes of the same
      column.  Note that $\sum (2S_{\text{tot}}+1)=2^N$.}
    \label{fig:hsfoursitesu2}
  \end{center}
\end{figure*}
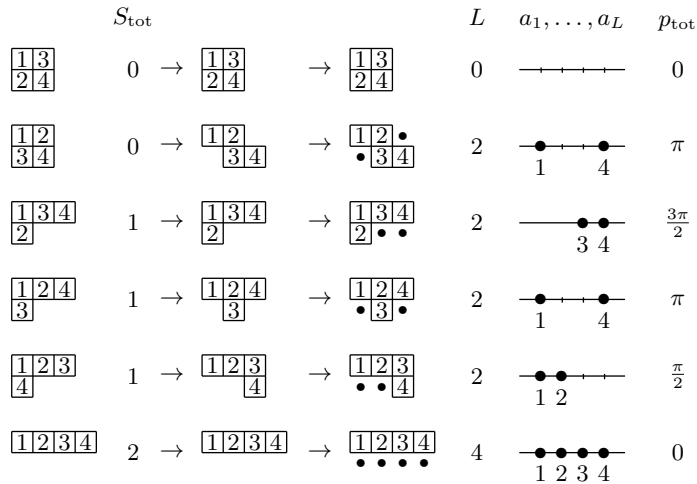

The one-to-one correspondence between the Young tableaux and the
non-interacting many-spinon eigenstates of the Haldane--Shastry model
is illustrated in Fig.~\ref{fig:hsfoursitesu2} for a chain with $N=4$
sites.  The rule is that in each Young tableau, we shift boxes to the
right such that each box is below or in the column to the right of the
box with the preceding number.  Each missing box in the resulting,
extended tableaux represents a spinon.  The extended tableaux provide
us with the total spin of each multiplet, which is given by the
representation specified by the original Young tableau, the
number $L$ of spinons present, and the individual spinon momentum
numbers $a_i$, which are just the numbers in the boxes above or below
the dots representing the spinons.  The single-spinon momenta are
obtained from those via
\begin{align}
  \label{eq:hssinglespinonmom}
  p_i=\frac{\pi}{N}\,\left(a_i-\frac{1}{2}\right),
\end{align}
which implies $\delta\le p_i\le\pi-\delta$, with
$\delta=\frac{\pi}{2N}\to 0$ for $N\to\infty$.

The total momentum and the total energies of the many-spinon
states are given by \eqref{eq:hsLspinonenergy} with
\begin{align}
  \label{eq:hssutwopzeroezero}
  p_0=-\frac{\pi}{2}\:N,\quad 
  E_0=-\frac{\pi^2}{24}\left(N+\frac{5}{N}\right),
\end{align}
and the single-spinon dispersion
\begin{align}
  \label{eq:hssutwoepsilon}
  \epsilon(p)=\frac{1}{2}p\left(\pi-p\right)
    +\frac{\pi^2}{8N^2},
\end{align}
where we use a convention according to which the ``vacuum'' state
$\ket{\dw\dw\ldots\dw}$ has momentum $p=0$ (and the empty state
$\ket{0}$ has $p=\pi (N-1)$).

\vspace{.5\baselineskip} This correspondence shows that spinons are
non-interacting, with momentum spacings appropriate for half-fermions.
We may interpret the Haldane-Shastry model as a reparameterization of
a Hilbert space spanned by spin flips (\ref{eq:hsket}) into a basis
which consists of the Haldane-Shastry ground state plus all possible
many-spinon states.  



\section{Majorana spacings and Ising anyons} 
\label{sec:na}

\subsection{An exact model of a critcal $S=1$ spin chain described 
by a Pfaffian}
\label{sec:namod}

As for the Haldane--Shastry model, we consider a one-dimensional
lattice with periodic boundary conditions and an even number of sites
$N$ on a unit circle embedded in the complex plane.  The only
difference is that now the spin on each site is $S=1$:
\begin{center}
\begin{picture}(310,45)(-30,-25)
\put(0,0){\circle{100}}
\put( 20.0,   .0){\circle*{3}}
\put( 17.3, 10.0){\circle*{3}}
\put( 10.0, 17.3){\circle*{3}}
\put(   .0, 20.0){\circle*{3}}
\put(-10.0, 17.3){\circle*{3}}
\put(-17.3, 10.0){\circle*{3}}
\put(-20.0,   .0){\circle*{3}}
\put(-17.3,-10.0){\circle*{3}}
\put(-10.0,-17.3){\circle*{3}}
\put(   .0,-20.0){\circle*{3}}
\put( 10.0,-17.3){\circle*{3}}
\put( 17.3,-10.0){\circle*{3}}
\qbezier[20]( 20.0,   .0)(  5.0,8.65)(-10.0, 17.3)
\put(50,8){\makebox(0,0)[l]{$N$\ sites with spin 1 on unit circle:}}
\put(50,-8){\makebox(0,0)[l]{$\displaystyle \eta_\alpha=e^{\i \frac{2\pi}{N}\alpha }$
\ \ with\ $\alpha = 1,\ldots ,N$}}
\end{picture}
\end{center}

The ground state wave function we consider
here~\cite{greiter02jltp1029,
greiter-09prl207203,Greiter11
}
is given by a bosonic Pfaffian state in the complex lattice
coordinates $z_i$ supplemented by a phase factor,
\begin{align}
  \label{eq:napsi0}
  \psi^{S=1}_0(z_1,z_2,\ldots ,z_N) 
  =\text{Pf}\left(\frac{1}{z_{i}-z_{j}}\right)
  \prod_{i<j}^{N}(z_i-z_j)\prod_{i=1}^{N}\,z_i.
\end{align}
The Pfaffian is given by the fully antisymmetrized sum over all
possible pairings of the $N$ particle coordinates,
\begin{align}
  \label{eq:napfaff}
  \text{Pf}\left(\frac{1}{z_i -z_j}\right)\equiv
  \mathcal{A}
  \left\{
    \frac{1}{z_1-z_2}\cdot\,\ldots\,\cdot\frac{1}{z_{N-1}-z_{N}}
  \right\}.
\end{align}
The ``particles'' $z_i$ represent re-normalized spin flips
$\tilde{S}_{\alpha}^{+}$ acting on a vacuum with all spins in the
$S^{z}=-1$ state,
\begin{align}
  \label{eq:naket}
  \ket{\psi^{S=1}_0}=\sum_{\{z_1,\dots,z_{N}\}} 
  \psi^{S=1}_0(z_1,\dots,z_N)\
  \tilde{S}_{z_1}^{+}\cdot\dots\cdot\tilde{S}_{z_{N}}^{+} 
  \ket{-1}_N,
\end{align}
where the sum extends over all possibilities of distributing the 
$N$ ``particles'' over the $N$ lattice sites allowing for double 
occupation, 
\begin{align}
  \label{eq:naspinflip}
  \tilde{S}_{\alpha}^{+} \equiv\frac{{S}^{\rm{z}}_{\alpha}+1}{2} S_\alpha^{+},
\end{align}
and
\begin{align}
  \label{eq:navacuumket}
  \ket{-1}_N\equiv\otimes_{\alpha=1}^N \ket{1,-1}_{\alpha}.
\end{align}
This state is translationally invariant with momentum $p_0=0$, a spin
singlet, real, and invariant under P and T.  It may be viewed as the
one-dimensional analog of the non-Abelian chiral spin
liquid~\cite{greiter-09prl207203,greiter-14prb165125}.

Like the ground state of the Haldane--Shastry model, the $S=1$ state
\eqref{eq:napsi0} describes a critical spin liquid in one dimension,
with algebraically decaying correlations.  It does not, however, serve
as a paradigm of the generic $S=1$ spin state, as the generic state
possesses a Haldane
gap~\cite{haldane83pl464,haldane83prl1153,affleck90proc,Fradkin91} in
the spin excitation spectrum due to linearly confining forces between
the
spinons~\cite{affleck-87prl799,affleck-88cmp477,greiter02jltp1029,greiter-07prb184441,greiter10np5}.

The $S=1$ Pfaffian state \eqref{eq:naket} with
\eqref{eq:napsi0} is the exact ground state of the Hamiltonian~\cite{Greiter11,thomale-12prb195149}
  \begin{align}
    \label{eq:naham}
    H^{S=1}=&\frac{2\pi^2}{N^2} \Bigg[ \sum_{\substack{\a\ne
    \b}}^N\frac{\bSa\bSb}{\vert\ea-\eb\vert^2}
    \nonumber\\[5pt] &
    -\frac{1}{20}\sum_{\substack{\a,\b,\c\atop \a\ne\b,\c}}^N
    \frac{(\bSa\bSb)(\bSa\bSc) +
    (\bSa\bSc)(\bSa\bSb)}{(\eab-\ebb)(\ea-\ec)}
    \Bigg],\nonumber\\[-4pt]
  \end{align}
with energy eigenvalue 
\begin{align}
  \label{eq:b:e0}
  E_0^{S=1}
  =-\frac{2\pi^2}{15}\left(N+\frac{5}{N}\right).
\end{align}
This model was shortly afterwards independently rediscovered by
Nielsen, Cirac, and Sierra~\cite{nielsen-11jsmte11014}.  
The effective field theory of the Hamiltonian \eqref{eq:naham} is
given by the SU(2) level $k=2$ Wess--Zumino--Witten
model~\cite{wess-71plb95,witten84cmp455,affleck86npb409,affleck-87prb5291}.
It was further shown very recently by Michaud
\etal~\cite{michaud-13prb140404} that one can construct a critical
spin model of the Wess--Zumino--Witten universality class if one only
keeps the leading two- and three-body spin terms in \eqref{eq:naham},
and adjusts the coefficients accordingly.

In analogy to the non-Abelian quasiparticles of the Pfaffian state in
the quantized Hall effect, the spinons of the model are Ising anyons.
The space of the (in the thermodynamic limit) degenerate states
associated with the non-Abelian statistics is spanned by the Majorana
fermion orbitals at the quasiparticle or spinon excitations.  The
model hence provides us with a framework to study Ising anyons in one
dimension.

\subsection{Generation of the ground state by projection from Gutzwiller states}
\label{sec:napro}

We will show now that the $S=1$ ground state \eqref{eq:napsi0} can
alternatively be generated by considering two (identical)
Haldane--Shastry or Gutzwiller states \eqref{eq:hspsi0} and 
projecting onto the triplet or $S=1$ configuration 
contained in
\begin{align}
  \label{eq:nahalftimeshalf}
  \textstyle \bs{\frac{1}{2}}\otimes\bs{\frac{1}{2}} = \bs{0}\oplus \bs{1}
\end{align}
at each site~\cite{greiter02jltp1029,greiter-09prl207203}.  

This projection can be accomplished very conveniently using Schwinger
bosons~\cite{schwinger65proc,Auerbach94}.  In terms of those, the
SU(2) spin operators are given by
\begin{align}
  \label{eq:naSschwb}
  \bs{S}=\frac{1}{2}\left(a^\dagger,b^\dagger\right)\,\bs{\sigma}\!\left(\!\!
    \begin{array}{c}
      \,a\, \\[2pt] b
    \end{array}\!\!\right)\!,
\end{align}
where $\bs{\sigma}=(\sigma_\x,\sigma_\y,\sigma_\z )$ is the vector
consisting of the three Pauli matrices, and $a^\dagger,b^\dagger$
($a,b$) are independent boson creation (annihilation) operators which
obey 
\begin{align}
  \begin{array}{c}
    \comm{a}{a^\dagger}=\comm{b}{b^\dagger}=1,\\
    \comm{a^{\phantom{\dagger}}\!\!}{b}=\comm{a}{b^\dagger}
    =\comm{a^\dagger}{b}=\comm{a^\dagger}{b^\dagger}
    =0\rule{0pt}{16pt}.
  \end{array}
  \label{eq:schwb}
\end{align}
The spin quantum number $S$ is given by half the number of bosons,  
\begin{align}
2S=a^\dagger a + b^\dagger b,
\label{eq:schwt}
\end{align}
and the usual spin states (simultaneous eigenstates of $\boldsymbol{S}^2$
and $S^\z $) are given by
\begin{align}
\ket{S,m} = \frac{(a^\dagger)^{S+m}}{\sqrt{(S+m)!}} \frac{(b^\dagger)^{S-m}}
{\sqrt{(S-m)!}} \vac.
\label{eq:schs}
\end{align}
In particular, the spin-\half states are given by
\begin{align}
\ket{\up}=c_{\up}^\dagger \vac =a^\dagger \vac ,\qquad 
\ket{\dw}=c_{\dw}^\dagger \vac =b^\dagger \vac ,
\label{eq:schwfun}
\end{align}
\ie $a^\dagger$ and $b^\dagger$ act just like the fermion creation
operators $c^\dagger_\up$ and $c^\dagger_\dw$ in this case.  The
difference between fermion creation and Schwinger boson creation
operators shows up only when two (or more) creation operators act on
the same site or orbital.  The fermion operators create an
antisymmetric (or singlet) configuration (in accordance with the Pauli
principle),
\begin{align}
\ket{0,0} = c_{\up}^\dagger c_{\dw}^\dagger \vac ,
\label{eq:schwfer}
\end{align}
while the Schwinger bosons create a totally symmetric (triplet, or
higher spin if we create more than two bosons) configuration,
\begin{align}
  \label{eq:nas=1schw}
  \ket{1,1} &= \textstyle{\frac{1}{\sqrt{2}}} (a^\dagger)^{2}\vac 
  ,\nonumber\\[0.2\baselineskip]
  \ket{1,0} &= a^\dagger b^\dagger \vac , \\[0.2\baselineskip]
  \ket{1,-1} &= \textstyle{\frac{1}{\sqrt{2}}} (b^\dagger)^{2}\vac 
  .\nonumber
\end{align}

To generate the $S=1$ ground state \eqref{eq:napsi0}, we first rewrite
\eqref{eq:hsket} in terms of Schwinger bosons,
\begin{align}
  \label{eq:nahsket}
  \ket{\psi^{\s\text{HS}}_{0}}
  &=\sum_{\{z_1,\ldots ,z_M\}}
  \psi^{\s\text{HS}}_{0}[z]\,
  {S}^+_{z_1}\cdot\ldots\cdot {S}^+_{z_M} 
  \ket{\dw\dw\ldots\dw} 
  \nonumber\\[4pt]
  &=\hspace{-4pt}\sum_{\{z_1,\ldots ,z_M;w_1,\ldots,w_M\}}\hspace{-4pt}
  \psi^{\s\text{HS}}_{0}[z]\,
    {a}^+_{z_1}\ldots a^\dagger_{z_M}
    {b}^+_{w_1}\ldots b^\dagger_{w_M}%
  \vac\!\nonumber\\[6pt]
  &\equiv \Psi^{\s\text{HS}}_{0}[a^\dagger,b^\dagger] \vac\!,
\end{align}
where $M=\frac{N}{2}$ and the $w_k$'s are those lattice sites which
are not occupied by any of the $z_i$'s.  Then the Pfaffian state
\eqref{eq:naket} with \eqref{eq:napsi0} is up to an overall
normalization factor given by
\begin{align}
  \label{eq:naop}
  \ket{\psi^{S=1}_0}=
  \Big(\Psi^{\s\text{HS}}_0\big[a^\dagger ,b^\dagger\big]\Big)^2\vac.
\end{align}

To verify \eqref{eq:naop}, use the identity 
\begin{align}
  \label{eq:na220=Pf} 
  \mathcal{S}\Biggl\{
  \prod_{\substack{i,j=1\atop i<j}}^{M}\hspace{0pt}(z_i-z_j)^2\hspace{-2pt}
  \prod_{\substack{i,j=M+1\atop i<j}}^{2M}\hspace{-2pt}(z_i-z_j)^2\Biggr\}&
  \nonumber\\
  =\text{Pf}\left(\frac{1}{z_i-z_j}\right)
  \prod_{i<j}^{2M}(z_i-z_j),&
\end{align}
where $\mathcal{S}$ indicates symmetrization over all the variables in
the curly brackets, and
\begin{align}
  \label{eq:naStilde^n}
  \frac{1}{\sqrt{2}}(a^\dagger)^n (b^\dagger)^{(2-n)}\vac 
  =(\tilde{S}^+)^n\ket{1,-1},
\end{align}
which is readily verified with \eqref{eq:naSschwb},
\eqref{eq:nas=1schw}, and the definition \eqref{eq:naspinflip}.
To prove \eqref{eq:na220=Pf}, use the following identity due to
Frobenius~\cite{frobenius-1882ram53},
\begin{align}
  \label{eq:nacauchy}
  &\det\left(\frac{1}{z_i-z_{M+j}}\right)
  \nonumber\\
  &=(-1)^{\textstyle\frac{M(M+1)}{2}}\,  
  \frac{\displaystyle 
      \prod_{\substack{i,j=1\atop i<j}}^{M}(z_i-z_j)
      \prod_{\substack{i,j=M+1\atop i<j}}^{2M}(z_i-z_j)
    }{\displaystyle \prod_{i=1}^M\,\prod_{j=M+1}^{2M}(z_i-z_j)}.
\end{align}

The projective construction directly reveals that the state
\eqref{eq:naket} with \eqref{eq:napsi0} inherits several symmetries
from the Gutzwiller state 
(translational, SU(2) spin rotation, parity, and time reversal).

\subsection{Topological degeneracies and non-Abelian statistics}
\label{sec:nana}

It is well established that $2n$ spatially well separated
quasiparticle excitations or vortices carrying half of a Dirac flux
quanta each in the non-Abelian quantized Hall state described by the
Pfaffian~\cite{moore-91npb362,greiter-91prl3205,greiter-92npb567}
will span an internal or topological Hilbert space of dimensions $2^n$
($2^{n-1}$ for either even or odd fermion
numbers)~\cite{nayak-96npb529}, in accordance with the existence of
one Majorana fermion state at each vortex
core~\cite{kopnin-91prb9667,read-00prb10267,ivanov01prl268,stern-04prb205338}.
The Majorana fermion states can only be manipulated through braiding
of the vortices, with the interchanges being non-commutative or
non-Abelian~\cite{moore-91npb362,ivanov01prl268,stern10n187}.



The question we wish to address in this paper is whether there is
any manifestation of this topological space of dimension $2^n$, or the
$2n$ Majorana fermion states, in the spinon excitation Hilbert space
suggested by the $S=1$ ground state \eqref{eq:napsi0}.  In Section
\ref{sec:hsspinons}, we have seen that the fractional statistics of
the spinons in the Haldane--Shastry model, and presumably in any model
supporting one-dimensional anyons, is encoded in the momentum spacings
of the excitations.  This is not too surprising, as there are no other
suitable quantum numbers, like the relative angular momentum for
two-dimensional anyons, available.  We will propose now that the
topological degeneracies, or the occupation numbers of the $n$
fermions consisting of the $2n$ Majorana fermions, are once again
encoded in the momentum spacings between single spinon states.

In the Haldane--Shastry model, the spacings between neighboring
momenta were always half integer, in accordance with half-Fermi
statistics, as the difference between consecutive spinon momentum
numbers  $a_i$ was always an odd integer,
\begin{align}
  \label{eq:nahsspacings}
  a_{i+1}-a_{i}\,=\,\text{odd}.
\end{align}
This follows directly from the construction of the extended Young
tableaux illustrated in Fig.\ \ref{fig:hsfoursitesu2}.  When two spinons
are in neighboring columns, the difference of the $a_i$ is one and
hence an odd integer; when we insert complete columns without spinons in
between, the number of boxes we insert is always even.

\newcommand{\yt}[3]{\put(#1,#2){\framebox(0.94,0.94){#3}}}
\newcommand{\ye}[3]{\put(#1,#2){\makebox(0.94,0.94){#3}}}
\newcommand{\yd}[2]{\put(#1,#2){\makebox(1.38,0.92){\circle*{0.4}}}}
\begin{figure*}[tb]
  \begin{center}
    \setlength{\unitlength}{\ytlength}
    \begin{picture}(0.3,0)(-1,10.9) \linethickness{0.5pt}
      \put(0,0){\ye{-0.5}{14.5}{$=$}
      \yt{1}{15}{1}
      \yt{2}{15}{1}
      \yt{1}{14}{2}
      \yt{2}{14}{2}
      \ye{3.5}{14.5}{$\oplus$}
      \yt{5}{15}{1}
      \yd{5}{14}
      \yt{6}{15}{1}
      \yt{6}{14}{2}
      \yt{7}{15}{2}
      \yd{7}{14}
      \put(0,0){\ye{1.5}{11.5}{$\oplus$}
      \yt{3}{12}{1}
      \yd{3}{11}
      \yt{4}{12}{1}
      \yd{4}{11}
      \yt{5}{12}{2}
      \yd{5}{11}
      \yt{6}{12}{2}
      \yd{6}{11}}}
    \end{picture}
    \begin{picture}(58,9)(1,10.5) \linethickness{0.5pt}
      \yt{1}{18}{1}
      \yt{2}{18}{1}
      \ye{3.5}{18}{$\otimes$}
      \yt{5}{18}{2}
      \yt{6}{18}{2}
      \ye{7.5}{18}{$\otimes$}
      \yt{9}{18}{3}
      \yt{10}{18}{3}
      \ye{12}{18}{$=$}
      \put(4,17){\makebox(0,0){$\underbrace{\phantom{\hspace{54pt}}}$}}
      \yt{14}{18}{1}
      \yt{15}{18}{1}
      \yt{14}{17}{2}
      \yt{15}{17}{2}
      \yt{16}{18}{3}
      \yt{17}{18}{3}
      \yd{16}{17}
      \yd{17}{17}
      \ye{15.5}{15.3}{$S=1$}
      \ye{18.5}{17.5}{$\oplus$}
      \yt{20}{18}{1}
      \yd{20}{17}
      \yt{21}{18}{1}
      \yt{21}{17}{2}
      \yt{22}{18}{2}
      \yt{22}{17}{3}
      \yt{23}{17}{3}
      \yd{23}{18}
      \ye{21.5}{15.3}{$S=0$}
      \ye{24.5}{17.5}{$\oplus$}
      \yt{26}{18}{1}
      \yd{26}{17}
      \yt{27}{18}{1}
      \yt{27}{17}{2}
      \yt{28}{18}{2}
      \yt{28}{17}{3}
      \yt{29}{18}{3}
      \yd{29}{17}
      \ye{27.5}{15.3}{$S=1$}
      \ye{30.5}{17.5}{$\oplus$}
      \yt{32}{18}{1}
      \yd{32}{17}
      \yt{33}{18}{1}
      \yt{33}{17}{2}
      \yt{34}{18}{2}
      \yd{34}{17}
      \yt{35}{18}{3}
      \yd{35}{17}
      \yt{36}{18}{3}
      \yd{36}{17}
      \ye{34}{15.3}{$S=2$}
      \put(23,6){
      \ye{14.5}{11.5}{$\oplus$}
      \yt{16}{12}{1}
      \yd{16}{11}
      \yt{17}{12}{1}
      \yd{17}{11}
      \yt{18}{12}{2}
      \yt{18}{11}{3}
      \yt{19}{12}{2}
      \yt{19}{11}{3}
      \ye{17.5}{9.3}{$S=1$}
      \ye{20.5}{11.5}{$\oplus$}
      \yt{22}{12}{1}
      \yd{22}{11}
      \yt{23}{12}{1}
      \yd{23}{11}
      \yt{24}{12}{2}
      \yd{24}{11}
      \yt{25}{12}{2}
      \yt{25}{11}{3}
      \yt{26}{12}{3}
      \yd{26}{11}
      \ye{24}{9.3}{$S=2$}
      \ye{27.5}{11.5}{$\oplus$}
      \yt{29}{12}{1}
      \yd{29}{11}
      \yt{30}{12}{1}
      \yd{30}{11}
      \yt{31}{12}{2}
      \yd{31}{11}
      \yt{32}{12}{2}
      \yd{32}{11}
      \yt{33}{12}{3}
      \yd{33}{11}
      \yt{34}{12}{3}
      \yd{34}{11}
      \ye{31.5}{9.3}{$S=3$}}
    \end{picture}
    \caption{Total spin representations of three $S=1$ spins
      in terms of extended Young tableaux.}
    \label{fig:nayoungdiagram}
  \end{center}
\end{figure*}


We will now show that for the $S=1$ chain with the Hilbert space
parameterized by the ground state $\ket{\psi^{S=1}_0}$ and all spinon
excitations above it, the corresponding rule is
\begin{align}
   \label{eq:naspacings}
   \begin{array}{rcl@{\hspace{20pt}}l}
     a_{i+1}-a_{i}&=&\text{even\ or\ odd,} &\text{for}\ i\ \text{odd},
     \\[2pt]
     a_{i+1}-a_{i}&=&\text{odd,}         &\text{for}\ i\ \text{even}.
   \end{array}
\end{align}
As $i=1,2,\ldots,2n$, we have a total of $n$ spacings which can be
either even or odd, and another $n$ spacings which are always odd.
With the single spinon momenta given by
\begin{align}
  \label{eq:nasinglespinonmom}
  p_i=\frac{\pi}{N}\,\left(a_i-\frac{1}{2}\right),
\end{align}
this yields momentum spacings which can be either an integer or
an half-integer times $\frac{2\pi}{N}$ for $i$ odd.  This is a
topological distinction---for Abelian anyons, one choice corresponds
to bosons or fermions (which, for most purposes, are equivalent in one
dimension), and the other choice to half-fermions.  For spinons which
are well separated in momentum space, the states spanning this in
total $2^n$ dimensional topological Hilbert space become degenerate as
we approach the thermodynamic limit.

\begin{figure*}[tb]
  \begin{center}
    \setlength{\unitlength}{\ytlength}
  \begin{picture}(21,3)(-1,3) \linethickness{0.5pt}
    \ye{8.5}{4}{$S_{\text{tot}}$}
    \ye{12.5}{4}{$a_1,\dots,a_L$}
    \ye{16.5}{4}{$p_{\text{tot}}$}
  \end{picture}
  \begin{picture}(21,3)(-1,3) \linethickness{0.5pt}
    \ye{8.5}{4}{$S_{\text{tot}}$}
    \ye{12.5}{4}{$a_1,\dots,a_L$}
    \ye{16.5}{4}{$p_{\text{tot}}$}
  \end{picture}
  \begin{picture}(21,3)(-1,3) \linethickness{0.5pt}
    \ye{8.5}{4}{$S_{\text{tot}}$}
    \ye{12.5}{4}{$a_1,\dots,a_L$}
    \ye{16.5}{4}{$p_{\text{tot}}$}
  \end{picture}
  \begin{picture}(21,3.5)(-1,0) \linethickness{0.5pt}
    \yt{1}{2}{1}
    \yt{1}{1}{2}
    \yt{2}{2}{1}
    \yt{2}{1}{2}
    \yt{3}{2}{3}
    \yt{3}{1}{4}
    \yt{4}{2}{3}
    \yt{4}{1}{4}
    \ye{8.5}{1.5}{$0$}
    \put(10.5,2){\line(1,0){5}} 
    \multiput(11.5,1.85)(1,0){4}{\rule{0.5pt}{2pt}}
    \ye{16.5}{1.5}{$0$}
  \end{picture}
    \begin{picture}(21,3.5)(-1,0) \linethickness{0.5pt}
      \yt{1}{2}{1}
      \yd{1}{1}
      \yt{2}{2}{1}
      \yt{2}{1}{2}
      \yt{3}{2}{2}
      \yt{3}{1}{3}
      \yd{4}{2}
      \yt{4}{1}{3}
      \yt{5}{2}{4}
      \yd{5}{1}
      \yt{6}{2}{4}
      \yd{6}{1}
      \ye{8.5}{1.5}{$1$}
      \put(10.5,2){\line(1,0){5}} 
      \multiput(11.5,1.85)(1,0){4}{\rule{0.5pt}{2pt}}
      \multiput(11.5,2)(1,0){1}{\circle*{0.5}}
      \multiput(13.5,2)(1,0){2}{\circle*{0.5}}
      \multiput(14.5,2.6)(3,0){1}{\circle*{0.5}}
      \ye{11}{0.5}{1}
      \ye{13}{0.5}{3}
      \ye{14}{0.5}{4}
      \ye{16.5}{1.5}{$\frac{\pi}{2}$}
    \end{picture}
    \begin{picture}(21,3.5)(-1,0) \linethickness{0.5pt}
      \yt{1}{2}{1}
      \yd{1}{1}
      \yt{2}{2}{1}
      \yt{2}{1}{2}
      \yt{3}{2}{2}
      \yd{3}{1}
      \yt{4}{2}{3}
      \yd{4}{1}
      \yt{5}{2}{3}
      \yt{5}{1}{4}
      \yt{6}{2}{4}
      \yd{6}{1}
      \ye{8.5}{1.5}{$2$}
      \put(10.5,2){\line(1,0){5}} 
      \multiput(11.5,1.85)(1,0){4}{\rule{0.5pt}{2pt}}
      \multiput(11.5,2)(1,0){4}{\circle*{0.5}}
      \ye{11}{0.5}{1}
      \ye{12}{0.5}{2}
      \ye{13}{0.5}{3}
      \ye{14}{0.5}{4}
      \ye{16.5}{1.5}{$0$}
    \end{picture}
    \begin{picture}(21,3.5)(-1,0) \linethickness{0.5pt}
      \yt{1}{2}{1}
      \yd{1}{1}
      \yt{2}{2}{1}
      \yt{2}{1}{2}
      \yt{3}{2}{2}
      \yt{3}{1}{3}
      \yt{4}{2}{3}
      \yt{4}{1}{4}
      \yd{5}{2}
      \yt{5}{1}{4}
      \ye{8.5}{1.5}{$0$}
      \put(10.5,2){\line(1,0){5}} 
      \multiput(11.5,1.85)(1,0){4}{\rule{0.5pt}{2pt}}
      \multiput(11.5,2)(3,0){2}{\circle*{0.5}}
      \ye{11}{0.5}{1}
      \ye{14}{0.5}{4}
      \ye{16.5}{1.5}{$\pi$}
    \end{picture}
    \begin{picture}(21,3.5)(-1,0) \linethickness{0.5pt}
      \yt{1}{2}{1}
      \yd{1}{1}
      \yt{2}{2}{1}
      \yd{2}{1}
      \yt{3}{2}{2}
      \yt{3}{1}{3}
      \yt{4}{2}{2}
      \yt{4}{1}{3}
      \yd{5}{2}
      \yt{5}{1}{4}
      \yt{6}{2}{4}
      \yd{6}{1}
      \ye{8.5}{1.5}{$1$}
      \put(10.5,2){\line(1,0){5}} 
      \multiput(11.5,1.85)(1,0){4}{\rule{0.5pt}{2pt}}
      \multiput(11.5,2)(3,0){2}{\circle*{0.5}}
      \multiput(11.5,2.6)(3,0){2}{\circle*{0.5}}
      \ye{11}{0.5}{1}
      \ye{14}{0.5}{4}
      \ye{16.5}{1.5}{$0$}
    \end{picture}
    \begin{picture}(21,3.5)(-1,0) \linethickness{0.5pt}
      \yt{1}{2}{1}
      \yd{1}{1}
      \yt{2}{2}{1}
      \yd{2}{1}
      \yt{3}{2}{2}
      \yd{3}{1}
      \yt{4}{2}{2}
      \yt{4}{1}{3}
      \yt{5}{2}{3}
      \yt{5}{1}{4}
      \yt{6}{2}{4}
      \yd{6}{1}
      \ye{8.5}{1.5}{$2$}
      \put(10.5,2){\line(1,0){5}}
      \multiput(11.5,1.85)(1,0){4}{\rule{0.5pt}{2pt}}
      \multiput(11.5,2)(1,0){2}{\circle*{0.5}}
      \multiput(11.5,2.6)(1,0){1}{\circle*{0.5}}
      \multiput(14.5,2)(1,0){1}{\circle*{0.5}}
      \put(11,0.5){\makebox(1,1){1}}
      \put(12,0.5){\makebox(1,1){2}}
      \put(14,0.5){\makebox(1,1){4}}
      \ye{16.5}{1.5}{$\frac{3\pi}{2}$}
    \end{picture}
    \begin{picture}(21,3.5)(-1,0) \linethickness{0.5pt}
      \yt{1}{2}{1}
      \yd{1}{1}
      \yt{2}{2}{1}
      \yd{2}{1}
      \yt{3}{2}{2}
      \yt{3}{1}{3}
      \yt{4}{2}{2}
      \yt{4}{1}{3}
      \yd{5}{2}
      \yt{5}{1}{4}
      \yd{6}{2}
      \yt{6}{1}{4}
      \ye{8.5}{1.5}{$0$}
      \put(10.5,2){\line(1,0){5}} 
      \multiput(11.5,1.85)(1,0){4}{\rule{0.5pt}{2pt}}
      \multiput(11.5,2)(3,0){2}{\circle*{0.5}}
      \multiput(11.5,2.6)(3,0){2}{\circle*{0.5}}
      \ye{11}{0.5}{1}
      \ye{14}{0.5}{4}
      \ye{16.5}{1.5}{$0$}
    \end{picture}
    \begin{picture}(21,3.5)(-1,0) \linethickness{0.5pt}
      \yt{1}{2}{1}
      \yd{1}{1}
      \yt{2}{2}{1}
      \yd{2}{1}
      \yt{3}{2}{2}
      \yd{3}{1}
      \yt{4}{2}{2}
      \yt{4}{1}{3}
      \yt{5}{2}{3}
      \yt{5}{1}{4}
      \yd{6}{2}
      \yt{6}{1}{4}
      \ye{8.5}{1.5}{$1$}
      \put(10.5,2){\line(1,0){5}}
      \multiput(11.5,1.85)(1,0){4}{\rule{0.5pt}{2pt}}
      \multiput(11.5,2)(1,0){2}{\circle*{0.5}}
      \multiput(11.5,2.6)(1,0){1}{\circle*{0.5}}
      \multiput(14.5,2)(1,0){1}{\circle*{0.5}}
      \put(11,0.5){\makebox(1,1){1}}
      \put(12,0.5){\makebox(1,1){2}}
      \put(14,0.5){\makebox(1,1){4}}
      \ye{16.5}{1.5}{$\frac{3\pi}{2}$}
    \end{picture}
    \begin{picture}(21,3.5)(-1,0) \linethickness{0.5pt}
      \yt{1}{2}{1}
      \yd{1}{1}
      \yt{2}{2}{1}
      \yd{2}{1}
      \yt{3}{2}{2}
      \yd{3}{1}
      \yt{4}{2}{2}
      \yd{4}{1}
      \yt{5}{2}{3}
      \yt{5}{1}{4}
      \yt{6}{2}{3}
      \yt{6}{1}{4}
      \ye{8.5}{1.5}{$2$}
      \put(10.5,2){\line(1,0){5}}
      \multiput(11.5,1.85)(1,0){4}{\rule{0.5pt}{2pt}}
      \multiput(11.5,2)(1,0){2}{\circle*{0.5}}
      \multiput(11.5,2.6)(1,0){2}{\circle*{0.5}}
      \put(11,0.5){\makebox(1,1){1}}
      \put(12,0.5){\makebox(1,1){2}}
      \ye{16.5}{1.5}{$\pi$}
    \end{picture}
    \begin{picture}(21,3.5)(-1,0) \linethickness{0.5pt}
      \yt{1}{2}{1}
      \yt{1}{1}{2}
      \yt{2}{2}{1}
      \yt{2}{1}{2}
      \yt{3}{2}{3}
      \yd{3}{1}
      \yt{4}{2}{3}
      \yt{4}{1}{4}
      \yt{5}{2}{4}
      \yd{5}{1}
      \ye{8.5}{1.5}{$1$}
      \put(10.5,2){\line(1,0){5}} 
      \multiput(11.5,1.85)(1,0){4}{\rule{0.5pt}{2pt}}
      \multiput(13.5,2)(1,0){2}{\circle*{0.5}}
      \ye{13}{0.5}{3}
      \ye{14}{0.5}{4}
      \ye{16.5}{1.5}{$\frac{3\pi}{2}$}
    \end{picture}
  \begin{picture}(21,3.5)(-1,0) \linethickness{0.5pt}
      \yt{1}{2}{1}
      \yt{1}{1}{2}
      \yt{2}{2}{1}
      \yt{2}{1}{2}
      \yt{3}{2}{3}
      \yd{3}{1}
      \yt{4}{2}{3}
      \yd{4}{1}
      \yt{5}{2}{4}
      \yd{5}{1}
      \yt{6}{2}{4}
      \yd{6}{1}
      \ye{8.5}{1.5}{$2$}
      \put(10.5,2){\line(1,0){5}} 
      \multiput(11.5,1.85)(1,0){4}{\rule{0.5pt}{2pt}}
      \multiput(13.5,2)(1,0){2}{\circle*{0.5}}
      \multiput(13.5,2.6)(1,0){2}{\circle*{0.5}}
      \ye{13}{0.5}{3}
      \ye{14}{0.5}{4}
      \ye{16.5}{1.5}{$\pi$}
    \end{picture}
    \begin{picture}(21,3.5)(0,0) \linethickness{0.5pt}
      \yt{1}{2}{1}
      \yd{1}{1}
      \yt{2}{2}{1}
      \yt{2}{1}{2}
      \yt{3}{2}{2}
      \yd{3}{1}
      \yt{4}{2}{3}
      \yd{4}{1}
      \yt{5}{2}{3}
      \yd{5}{1}
      \yt{6}{2}{4}
      \yd{6}{1}
      \yt{7}{2}{4}
      \yd{7}{1}
      \ye{9.5}{1.5}{$3$}
      \put(11.5,2){\line(1,0){5}} 
      \multiput(12.5,1.85)(1,0){4}{\rule{0.5pt}{2pt}}
      \multiput(12.5,2)(1,0){4}{\circle*{0.5}}
      \multiput(14.5,2.6)(1,0){2}{\circle*{0.5}}
      \ye{12}{0.5}{1}
      \ye{13}{0.5}{2}
      \ye{14}{0.5}{3}
      \ye{15}{0.5}{4}
      \ye{17.5}{1.5}{$\frac{3\pi}{2}$}
    \end{picture}
    \begin{picture}(21,3.5)(-1,0) \linethickness{0.5pt}
      \yt{1}{2}{1}
      \yd{1}{1}
      \yt{2}{2}{1}
      \yt{2}{1}{2}
      \yt{3}{2}{2}
      \yt{3}{1}{3}
      \yt{4}{2}{3}
      \yt{4}{1}{4}
      \yt{5}{2}{4}
      \yd{5}{1}
      \ye{8.5}{1.5}{$1$}
      \put(10.5,2){\line(1,0){5}} 
      \multiput(11.5,1.85)(1,0){4}{\rule{0.5pt}{2pt}}
      \multiput(11.5,2)(3,0){2}{\circle*{0.5}}
      \ye{11}{0.5}{1}
      \ye{14}{0.5}{4}
      \ye{16.5}{1.5}{$\pi$}
    \end{picture}
    \begin{picture}(21,3.5)(-1,0) \linethickness{0.5pt}
      \yt{1}{2}{1}
      \yd{1}{1}
      \yt{2}{2}{1}
      \yt{2}{1}{2}
      \yt{3}{2}{2}
      \yt{3}{1}{3}
      \yt{4}{2}{3}
      \yd{4}{1}
      \yt{5}{2}{4}
      \yd{5}{1}
      \yt{6}{2}{4}
      \yd{6}{1}
      \ye{8.5}{1.5}{$2$}
      \put(10.5,2){\line(1,0){5}} 
      \multiput(11.5,1.85)(1,0){4}{\rule{0.5pt}{2pt}}
      \multiput(11.5,2)(1,0){1}{\circle*{0.5}}
      \multiput(13.5,2)(1,0){2}{\circle*{0.5}}
      \multiput(14.5,2.6)(3,0){1}{\circle*{0.5}}
      \ye{11}{0.5}{1}
      \ye{13}{0.5}{3}
      \ye{14}{0.5}{4}
      \ye{16.5}{1.5}{$\frac{\pi}{2}$}
    \end{picture}
    \begin{picture}(21,3.5)(0,0) \linethickness{0.5pt}
      \yt{1}{2}{1}
      \yd{1}{1}
      \yt{2}{2}{1}
      \yd{2}{1}
      \yt{3}{2}{2}
      \yd{3}{1}
      \yt{4}{2}{2}
      \yt{4}{1}{3}
      \yt{5}{2}{3}
      \yd{5}{1}
      \yt{6}{2}{4}
      \yd{6}{1}
      \yt{7}{2}{4}
      \yd{7}{1}
      \ye{9.5}{1.5}{$3$}
      \put(11.5,2){\line(1,0){5}} 
      \multiput(12.5,1.85)(1,0){4}{\rule{0.5pt}{2pt}}
      \multiput(12.5,2)(1,0){4}{\circle*{0.5}}
      \multiput(12.5,2.6)(3,0){2}{\circle*{0.5}}
      \ye{12}{0.5}{1}
      \ye{13}{0.5}{2}
      \ye{14}{0.5}{3}
      \ye{15}{0.5}{4}
      \ye{17.5}{1.5}{$\pi$}
    \end{picture}
    \begin{picture}(21,3.5)(-1,0) \linethickness{0.5pt}
      \yt{1}{2}{1}
      \yd{1}{1}
      \yt{2}{2}{1}
      \yt{2}{1}{2}
      \yt{3}{2}{2}
      \yd{3}{1}
      \yt{4}{2}{3}
      \yt{4}{1}{4}
      \yt{5}{2}{3}
      \yt{5}{1}{4}
      \ye{8.5}{1.5}{$1$}
      \put(10.5,2){\line(1,0){5}} 
      \multiput(11.5,1.85)(1,0){4}{\rule{0.5pt}{2pt}}
      \multiput(11.5,2)(1,0){2}{\circle*{0.5}}
      \ye{11}{0.5}{1}
      \ye{12}{0.5}{2}
      \ye{13}{0.5}{3}
      \ye{14}{0.5}{4}
      \ye{16.5}{1.5}{$\frac{\pi}{2}$}
    \end{picture}
    \begin{picture}(21,3.5)(-1,0) \linethickness{0.5pt}
      \yt{1}{2}{1}
      \yd{1}{1}
      \yt{2}{2}{1}
      \yd{2}{1}
      \yt{3}{2}{2}
      \yt{3}{1}{3}
      \yt{4}{2}{2}
      \yt{4}{1}{3}
      \yt{5}{2}{4}
      \yd{5}{1}
      \yt{6}{2}{4}
      \yd{6}{1}
      \ye{8.5}{1.5}{$2$}
      \put(10.5,2){\line(1,0){5}} 
      \multiput(11.5,1.85)(1,0){4}{\rule{0.5pt}{2pt}}
      \multiput(11.5,2)(3,0){2}{\circle*{0.5}}
      \multiput(11.5,2.6)(3,0){2}{\circle*{0.5}}
      \ye{11}{0.5}{1}
      \ye{14}{0.5}{4}
      \ye{16.5}{1.5}{$0$}
    \end{picture}
    \begin{picture}(21,3.5)(0,0) \linethickness{0.5pt}
      \yt{1}{2}{1}
      \yd{1}{1}
      \yt{2}{2}{1}
      \yd{2}{1}
      \yt{3}{2}{2}
      \yd{3}{1}
      \yt{4}{2}{2}
      \yd{4}{1}
      \yt{5}{2}{3}
      \yd{5}{1}
      \yt{6}{2}{3}
      \yt{6}{1}{4}
      \yt{7}{2}{4}
      \yd{7}{1}
      \ye{9.5}{1.5}{$3$}
      \put(11.5,2){\line(1,0){5}} 
      \multiput(12.5,1.85)(1,0){4}{\rule{0.5pt}{2pt}}
      \multiput(12.5,2)(1,0){4}{\circle*{0.5}}
      \multiput(12.5,2.6)(1,0){2}{\circle*{0.5}}
      \ye{12}{0.5}{1}
      \ye{13}{0.5}{2}
      \ye{14}{0.5}{3}
      \ye{15}{0.5}{4}
      \ye{17.5}{1.5}{$\frac{\pi}{2}$}
    \end{picture}
    \begin{picture}(21,3)(1,0.5) \linethickness{0.5pt}
      \yt{1}{2}{1}
      \yd{1}{1}
      \yt{2}{2}{1}
      \yd{2}{1}
      \yt{3}{2}{2}
      \yd{3}{1}
      \yt{4}{2}{2}
      \yd{4}{1}
      \yt{5}{2}{3}
      \yd{5}{1}
      \yt{6}{2}{3}
      \yd{6}{1}
      \yt{7}{2}{4}
      \yd{7}{1}
      \yt{8}{2}{4}
      \yd{8}{1}
      \ye{10.5}{1.5}{$4$}
      \put(12.5,2){\line(1,0){5}} 
      \multiput(13.5,1.85)(1,0){4}{\rule{0.5pt}{2pt}}
      \multiput(13.5,2)(1,0){4}{\circle*{0.5}}
      \multiput(13.5,2.6)(1,0){4}{\circle*{0.5}}
      \ye{13}{0.5}{1}
      \ye{14}{0.5}{2}
      \ye{15}{0.5}{3}
      \ye{16}{0.5}{4}
      \ye{18.5}{1.5}{$0$}
    \end{picture}
    \caption{Extended Young tableau decomposition 
      for an $S=1$ spin chain with $N=4$ sites.  The dots represent
      the spinons.  The spinon momentum numbers $a_i$ are given by the
      numbers in the boxes of the same column.  Note that $\sum
      (2S_{\text{tot}}+1)=3^N$.}
    \label{fig:nafoursitesu2}
  \end{center}
\end{figure*}
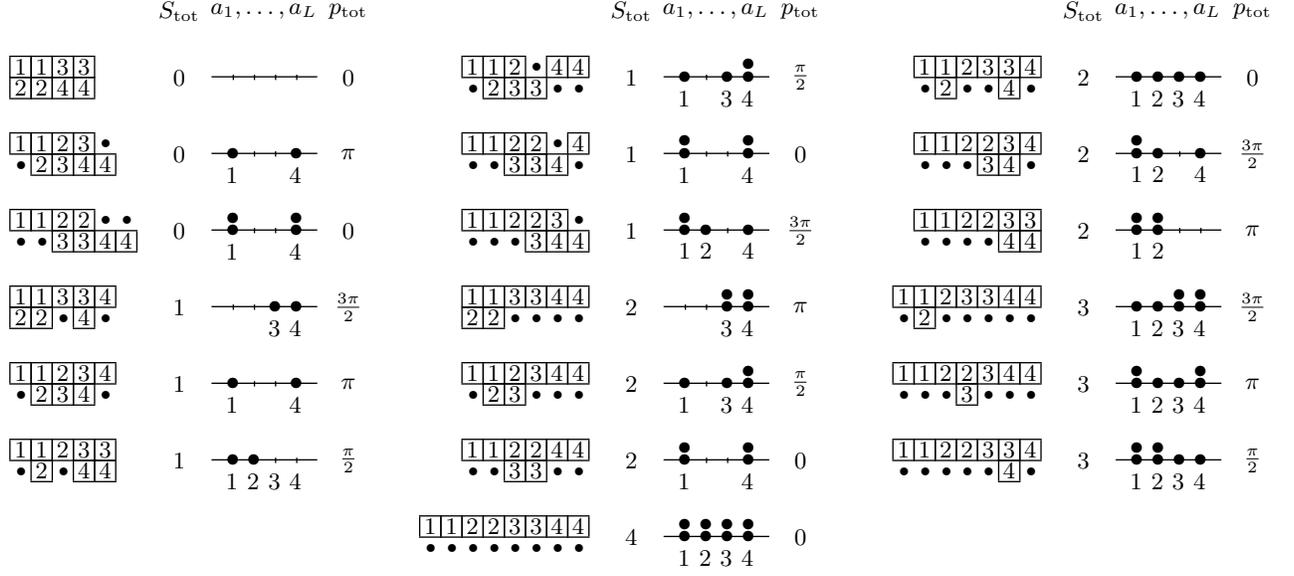

 
To derive \eqref{eq:naspacings}, we introduce a second formalism of
extended Young tableaux, 
this time for spin $S=1$.  The general rule we wish to propose for
obtaining the tableaux is illustrated in Fig.~\ref{fig:nayoungdiagram}
for three spins with $S=1$.  The construction is as follows.
For each of the $N$ spins, put a row of two adjacent boxes, which is
equivalent to the Young tableau for a single spin without any numbers
in the boxes.  Put these $N$ small tableaux on a line and number them
consecutively from left to right, with the same number in each pair of
boxes which represent a single spin.  To obtain the product of some
extended Young tableau representing spin ${S}_0$ on the left with a spin
$1$ tableau (\ie a row of two boxes with the same number in it) on the
right, we follow the rule
\begin{align}
  \label{eq:nastimes1}
  \bs{S}_0\otimes \bs{1} = \left\{
      \begin{array}{ll} 
        \bs{1},                                          
        &\text{for}\ \, S_0=0, \\[2pt] 
        \bs{S}_0\bs{-1}\,\oplus\,\bs{S}_0\,\oplus\,\bs{S}_0\bs{+1},\quad 
        &\text{for}\ \, S_0=1,2,\ldots 
      \end{array}\right.
\end{align}
\ie we obtain only one new tableau with both boxes from the right
added to the top row if the tableau on the left is a singlet, and
three new tableaux if it has spin one or higher.  These three tableaux
are constructed by adding both boxes to the bottom row (resulting in a
representation $\bs{S}_0\bs{-1}$), by adding the first box to the
bottom row and the second box to the top row without stacking them on
top of each other (resulting in a representation $\bs{S}_0$), and by
adding both boxes to the top row (resulting in a representation
$\bs{S}_0\bs{+1}$).  In each extended tableau, the boxes must be
arranged such that the numbers are strictly increasing in each column
from top to bottom, and that they are not decreasing from left to
right in that the smallest number in each column cannot be smaller
than the largest number in the column to the left of it.  In analogy
to the case of the Haldane--Shastry model, the empty spaces in between
the boxes are filled with dots representing spinons.  The spinon
momentum number $a_i$ associated with each spinon is given by the
number in the box in the same column.  A complete table of all the
extended Young tableaux for four $S=1$ spins is shown in
Fig.~\ref{fig:nafoursitesu2}.  The assignment of physical single
spinon momenta to the spinon momentum numbers
\eqref{eq:nasinglespinonmom} is identical to this assignment for the
Haldane--Shastry model, as we can obtain the $3^N$ states of the $S=1$
Hilbert space by Schwinger boson projection (\ie by projecting on spin
$S=1$ on each site) from states contained in the $2^N\times 2^N$
dimensional Hilbert space of two $S=\frac{1}{2}$ models, a projection
which commutes with the total momentum.  The correctness of this
assignment has further been verified numerically up to $N=16$
sites~\cite{scharfenberger-12jpa455202}.

With the tableau structure thus in place, all that is left to show is
that the allowed momentum spacings obey
\eqref{eq:naspacings}.  Looking at any of the tableaux in
Fig.~\ref{fig:nafoursitesu2}, we note that from left to right, the
spinons alternate between being assigned to the first of the two boxes
with a given number and being assigned to the second of such two
boxes.  This follows simply form the fact that the number of boxes in
between the columns with the two neighboring spinons must be even.
The first spinon momentum number $a_1$ is always odd, but all the
other $a_i$'s can be either even or odd.  The rule is therefore that
if $i$ is odd, the $i$-th spinon is assigned to the first of the two
boxes with number $a_i$, and the momentum spacing $a_{i+1}-a_i$ can be
either even or odd,
\begin{center}
  \setlength{\unitlength}{\ytlength}
  \begin{picture}(21,4)(0,0.5) \linethickness{0.5pt}
      \yt{1}{3}{\small 3}
      \yd{1}{2}
      \yt{2}{3}{\small 3}
      \yd{2}{2}
      \ye{1.5}{0.5}{\small even}
      \ye{4}{2.5}{or}
      \yt{6}{3}{\small 3}
      \yd{6}{2}
      \yt{7}{3}{\small 3}
      \yt{7}{2}{\small 4}
      \yt{8}{3}{\small 4}
      \yd{8}{2}
      \ye{7}{0.5}{\small odd}
      \ye{10}{2.5}{or}
      \yt{12}{3}{\small 3}
      \yd{12}{2}
      \yt{13}{3}{\small 3}
      \yt{13}{2}{\small 4}
      \yt{14}{3}{\small 4}
      \yt{14}{2}{\small 5}
      \yt{15}{3}{\small 5}
      \yd{15}{2}
      \ye{13.5}{0.5}{\small even}
      \ye{17}{2.5}{or}
      \ye{19}{2.2}{\ldots}
   \end{picture}
\end{center}
If $i$ is even, however, the $i$-th spinon is assigned to the second of
the two boxes with number $a_i$, and the momentum spacing
$a_{i+1}-a_i$ has to be odd, as we can insert only an even number of
columns between the two spinons (recall that we cannot stack two boxes
with the same number in it on top of each other):
\begin{center}
  \setlength{\unitlength}{\ytlength}
  \begin{picture}(24,4)(0,0.5) \linethickness{0.5pt}
      \yt{1}{3}{\small 3}
      \yd{1}{2}
      \yt{2}{3}{\small 4}
      \yd{2}{2}
      \ye{1.5}{0.5}{\small odd}
      \ye{4}{2.5}{or}
      \yt{6}{3}{\small 3}
      \yd{6}{2}
      \yt{7}{3}{\small 4}
      \yt{7}{2}{\small 5}
      \yt{8}{3}{\small 4}
      \yt{8}{2}{\small 5}
      \yt{9}{3}{\small 6}
      \yd{9}{2}
      \ye{7.5}{0.5}{\small odd}
      \ye{11}{2.5}{or}
      \yt{13}{3}{\small 3}
      \yd{13}{2}
      \yt{14}{3}{\small 4}
      \yt{14}{2}{\small 5}
      \yt{15}{3}{\small 4}
      \yt{15}{2}{\small 5}
      \yt{16}{3}{\small 6}
      \yt{16}{2}{\small 7}
      \yt{17}{3}{\small 6}
      \yt{17}{2}{\small 7}
      \yt{18}{3}{\small 8}
      \yd{18}{2}
      \ye{15.5}{0.5}{\small odd}
      \ye{20}{2.5}{or} 
      \ye{22}{2.2}{\ldots}
   \end{picture}
\end{center}
The spacings between the single spinon momenta are hence as 
stated in \eqref{eq:naspacings}.




\section{Topological momentum spacings for SU(2) level $k$ anyons in
  general}
\label{sec:naa}

\subsection{Generalization of the model to arbitrary spin S}
\label{sec:naarbs}

The projective generation introduced in Section \ref{sec:napro} can be
generalized to arbitrary spin $S=s$: 
\begin{align}
  \label{eq:nacslopS}
   \ket{\psi^{S}_0}=
   \Big(\Psi^{\s\text{HS}}_0\big[a^\dagger ,b^\dagger\big]\Big)^{2s}\vac.
\end{align} 
In order to write this state in a form similar to
\eqref{eq:napsi0}--\eqref{eq:navacuumket},
\begin{align}
  \label{eq:naketS}
  \ket{\psi^S_0}\;=\sum_{\{z_1,\dots,z_{sN}\}} 
  \psi^S_0(z_1,\dots,z_{sN})\
  \tilde{S}_{z_1}^{+}\cdot\dots\cdot\tilde{S}_{z_{sN}}^{+} 
  \ket{-s}_N,
\end{align}
where 
\begin{align}
  \label{eq:navacuumketS}
  \ket{-s}_N\equiv\otimes_{\alpha=1}^N \ket{s,-s}_{\alpha}
\end{align}
is the ``vacuum'' state in which all the spins are maximally polarized
in the negative $\hat z$-direction, we introduce 
re-normalized spin flip operators $\tilde{S}^{+}$ which satisfy
\begin{align}
  \label{eq:naStilde^nS}
  \frac{1}{\sqrt{(2s)!}}(a^\dagger)^n (b^\dagger)^{(2s-n)}\vac 
  =(\tilde{S}^+)^n\ket{s,-s}.
\end{align}
If we assume a basis in which $S^\z $ is diagonal, we may write 
\begin{align}
  \label{eq:naspinflipS}
  \tilde{S}^{+} 
  \,\equiv\,\frac{1}{b^\dagger b+1}\,a^\dagger b 
  \,=\, \frac{1}{s-{S}^\z +1}\, S^{+}.
\end{align}
The wave function for the spin $S$ state \eqref{eq:nacslopS} is then
given by 
\begin{align}
  \label{eq:narr}
  \psi^S_0(z_1,\dots,z_{sN})
  =\prod_{m=1}^{2s}\left(
    \prod_{\substack{i,j=(m-1)M+1\atop i<j}}^{mM}(z_i-z_j)^2 
  \right)\,
  \prod_{i=1}^{sN}z_i.
\end{align}
where $M=\frac{N}{2}$.
Note that these states are similar to the Read-Rezayi
states~\cite{read-99prb8084} in the quantized Hall effect.

As for the $S=1$ state discussed in Section \ref{sec:napro},
the projective construction \eqref{eq:nacslopS} directly implies
several symmetries.  The state 
$\ket{\psi^{S}_{0}}$ is translationally invariant with ground state
momentum $p_0=-\pi NS$, a spin singlet, and real.

It was again shown by one of us~\cite{Greiter11} that
\eqref{eq:nacslopS} (or \eqref{eq:naketS} with \eqref{eq:narr}) is the
exact ground state of the Hamiltonian
\begin{widetext}
  \begin{align}
    \label{eq:c:h}
    H^{S}=\frac{2\pi^2}{N^2} \Bigg[ &\sum_{\substack{\a\ne\b}}
    \frac{\bSa\bSb}{\vert\ea-\eb\vert^2} -\frac{1}{2(s+1)(2s+3)}
    \sum_{\substack{\a,\b,\c\atop \a\ne\b,\c}}
    \frac{(\bSa\bSb)(\bSa\bSc) +
      (\bSa\bSc)(\bSa\bSb)}{(\eab-\ebb)(\ea-\ec)} \Bigg],
    \nonumber\\ 
  \end{align}
\end{widetext}
with energy eigenvalue
\begin{align}
  \label{eq:c:E_0}
  E_0^{S} 
  = -\frac{\pi^2}{6}\frac{s(s+1)^2}{2s+3}\left(N+\frac{5}{N}\right).
\end{align}
Note that the Haldane--Shastry Hamiltonian \eqref{eq:hsham} and the
$S=1$ Hamiltonian \eqref{eq:naham} are just special cases of this
general model.

\subsection{Momentum spacings and topological degeneracies}
\label{sec:na:MSforSpinS}

In Section \ref{sec:nana}, we have shown that the non-Abelian
statistics of the Pfaffian state \eqref{eq:napsi0}, and in particular
the topological degeneracies associated with the Majorana fermion
states, manifests itself in topological choices for
the (kinematic) momentum spacings of the spinon excitations above
the $S=1$ ground state \eqref{eq:napsi0}.  Specifically, we found that
if we label the single spinon momenta in ascending order by
$p_i<p_{i+1}$, the spacings $p_{i+1}-p_i$ can be either even or odd
multiples of $\frac{\pi}{N}$ if $i$ is odd, while it has to be an odd
multiple if $i$ is even.

 \newcommand{\ytt}[3]{\put(#1,#2){\framebox(1.94,0.94){#3}}}
 \newcommand{\yttt}[3]{\put(#1,#2){\framebox(2.94,0.94){#3}}}

 In this Section, we formulate the corresponding restrictions for the
 general spin $S$ chain with ground state \eqref{eq:nacslopS}.  We
 will first state the rules and then motivate them.  Recall that
 spinons are represented by dots placed in the empty spaces of
 extended Young tableaux, and that the momentum number $a_i$ of spinon
 $i$ is given by the number in the box it shares a column with.  For
 general spin $S$, the tableau describing the representation on each
 site is given by
\begin{equation*}
  \setlength{\unitlength}{\ytlength}
  \begin{picture}(8,3.5)(0,0.8) \linethickness{0.5pt}
    \yt{1}{3}{}
    \yt{2}{3}{}
    \yt{3}{3}{}
    \ytt{4}{3}{}
    \yt{6}{3}{}
    \ye{7}{2.5}{,}
    \put(4,2){\makebox(0,0){$\underbrace{\phantom{\hspace{52pt}}}$}}
    \put(4,1){\makebox(0,0){\small $2S$ boxes}}
   \end{picture}
\end{equation*}
\ie a horizontal array of $2S$ boxes indicating symmetrization, which
all contain the same number.  

\begin{figure*}[tb]
  \begin{center}
  \includegraphics[scale=1]{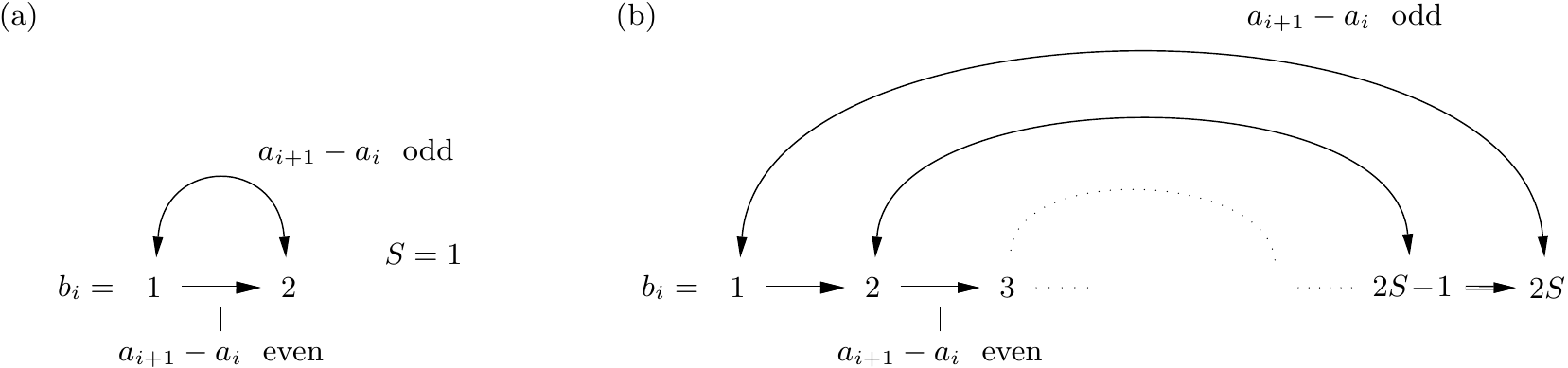}
  \end{center}
  \caption{Non-Abelian (SU(2) level $k=2S$) statistics in one
    dimension: flow diagram for the (auxiliary) box numbers $b_i$,
    which serve to describe the restrictions for the spinon momentum
    number spacings $a_{i+1}-a_i$ for the critical models of spin
    chains introduced in Sections \ref{sec:namod} and \ref{sec:naarbs}
    with (a) $S=1$, and (b) general spin $S$.  The unidirectional,
    horizontal arrows correspond to even integer momentum number
    spacings $a_{i+1}-a_i$, while the bidirectional, semicircle arrows
    correspond to odd integer spacings.}
  \label{fig:biflow}
\end{figure*}

If this number is $n$, the spinons we assign to any of these boxes
will have momentum number $a_i=n$.  Let us denote the number of the
box a given spinon $i$ with momentum number $a_i$ is assigned to, by
$b_i$, such that box number $b_i=1$ corresponds to the first, and box
number $b_i=2S$ to the last box with number $n$ in it:
\begin{equation*}
  \setlength{\unitlength}{\ytlength}
  \begin{picture}(8,5)(0,0.5) \linethickness{0.5pt}
    \yt{1}{3}{$n$}
    \yt{2}{3}{$n$}
    \yt{3}{3}{$n$}
    \ytt{4}{3}{}
    \yt{6}{3}{$n$}
    \ye{7}{2.5}{,}
    \yd{1}{2}
    \ye{1}{0.5}{\small $b_i=1$}
  \end{picture}
  \begin{picture}(10.5,5)(0,0.5) \linethickness{0.5pt}
    \yt{1}{3}{$n$}
    \yt{2}{3}{$n$}
    \yt{3}{3}{$n$}
    \ytt{4}{3}{}
    \yt{6}{3}{$n$}
    \ye{7}{2.5}{,}
    \ye{9}{2.5}{\ldots}
    \yd{2}{2}
    \ye{2}{0.5}{\small $b_i=2$}
   \end{picture}
  \begin{picture}(8,5)(0,0.5) \linethickness{0.5pt}
    \yt{1}{3}{$n$}
    \yt{2}{3}{$n$}
    \yt{3}{3}{$n$}
    \ytt{4}{3}{}
    \yt{6}{3}{$n$}
    \ye{7}{2.5}{.}
    \yd{6}{2}
    \ye{6}{0.5}{\small $b_i=2S$}
  \end{picture}
\end{equation*}
We will see below that if a representation of a spin $S$ chain
with $L$ spinons is written in terms of an extended Young tableau, the
first spinon with momentum number $a_1$ will always have box number
$b_1=1$, and the last spinon with $a_L$ will have $b_L=2S$.  The
restrictions corresponding to the non-Abelian (SU(2) level $k=2S$)
statistics of the spinons are described by the flow diagram of the
numbers $b_i$ shown in Figure \ref{fig:biflow}.


Let us elaborate this diagram first for the case $S=1$, which we
have already studied in Section \ref{sec:nana}.  In this case, 
\begin{align}
  \label{eq:nabi}
   b_i=\left\{
      \begin{array}{ll} 
        1,\quad &\text{for}\ \, i\ \text{odd}, \\[2pt] 
        2,\quad &\text{for}\ \, i\ \text{even}.
      \end{array}\right.
\end{align}
For $i$ odd, we may move from $b_i=1$ to $b_{i+1}=2$ either via the
horizontal arrow or via the semicircle in Figure \ref{fig:biflow}a,
and $a_{i+1}-a_i$ may hence be either even or odd, respectively.  For
$i$ even or $b_{i}=2$, however, the semicircle is the only
available continuation, which implies that the spacing $a_{i+1}-a_i$
must be odd.

For general $S$, Figure \ref{fig:biflow}b implies that the spacings
can be even or odd until $b_i=2S$ is reached, which is then followed
by an odd integer spacing $a_{i+1}-a_i$, as the semicircular arrow is
the only possible continuation at this point.  Note that for $S\ge 1$,
the minimal number of spinons is two (these two spinons then have an
odd integer spacing $a_2-a_1$), and that we cannot have more than $2S$
spinons with the same momentum number $a_i=n$, as $a_{i+1}-a_i=0$ is
even.

We will now motivate this diagram.  To begin with, we generalize the
formalism of extended Young tableaux to arbitrary spin $S$.  The
construction is similar to the one for $S=1$ outlined in Section
\ref{sec:nana}.  For each of the $N$ spins, put a row of $2S$ adjacent
boxes.  Put these $N$ tableaux on a line and number them consecutively
from left to right, with the same number in each row of $2S$ boxes
representing a single spin.  To obtain the product of some extended Young
tableau representing spin ${S}_0$ on the left with a spin $S$ tableau
(\ie a row of $2S$ boxes with the same number in it) on the right, we
first recall
\begin{align}
  \label{eq:naS0timesS}
  \bs{S}_0\otimes \bs{S} 
  = \,\vert\bs{S}_0\bs{-S}\vert\,
  \oplus\,\vert\bs{S}_0\bs{-S}\vert +1\,\oplus\,\ldots\,
  \oplus\,\,\bs{S}_0+\bs{S},
\end{align}
which implies that we obtain either $2S_0+1$ or $2S+1$ new tableaux,
depending on which number is smaller.  In terms of extended Young tableaux,
\eqref{eq:naS0timesS} translates into
\begin{align}
\label{eq:naS0timesSYT}
\includegraphics{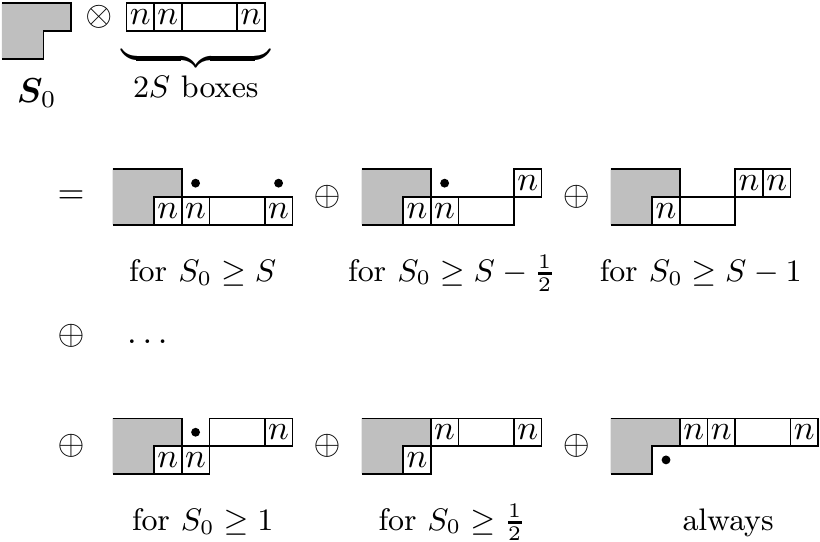}
\end{align}
The first tableau on the right-hand side of \eqref{eq:naS0timesSYT}
exists only for $S_0\ge S$, the second only for $S_0\ge
S-\frac{1}{2}$, and so on.  Note that the shape of the right boundary
of the extended Young tableaux for $\bs{S}_0$ does not determine which
tableaux are contained in the expansion of $\bs{S}_0\otimes \bs{S}$,
as this depends only on the number $S_0-S$.  In the expansion
\eqref{eq:naS0timesSYT}, the $2S$ boxes representing a single spin $S$
always reside in adjacent columns.  In an extended tableau, the
numbers in the boxes are equal or increasing as we go from left to
right, and strictly increasing from top to bottom.  The empty spaces
we obtain as we build up the tableaux via this method represent the
spinons.  Note that we cannot take a given tableau and just add a pair
of spinons by inserting them somewhere, as the resulting tableau would
not occur in the expansion.

\begin{figure*}[tb]
  \begin{center}
    \setlength{\unitlength}{\ytlength}
    \begin{picture}(52,5)(0,15) \linethickness{0.5pt}
       \yt{1}{18}{1}
       \yt{2}{18}{1}
       \yt{3}{18}{1}
       \yt{4}{18}{1}
     \ye{2.5}{16.3}{$S=2$}
     \ye{5.5}{18}{$\otimes$}
       \yt{7}{18}{2}
       \yt{8}{18}{2}
       \yt{9}{18}{2}
      \yt{10}{18}{2}
     \ye{8.5}{16.3}{$S=2$}

      \ye{12}{17.5}{$=$}
      \yt{14}{18}{1}
      \yt{14}{17}{2}
      \yt{15}{18}{1}
      \yt{15}{17}{2}
      \yt{16}{18}{1}
      \yt{16}{17}{2}
      \yt{17}{18}{1}
      \yt{17}{17}{2}
    \ye{15.5}{15.3}{$S=0$}

    \ye{18.5}{17.5}{$\oplus$}
      \yt{20}{18}{1}
      \yd{20}{17}
      \yt{21}{18}{1}
      \yt{21}{17}{2}
      \yt{22}{18}{1}
      \yt{22}{17}{2}
      \yt{23}{18}{1}
      \yt{23}{17}{2}
      \yt{24}{18}{2}
      \yd{24}{17}
      \ye{22}{15.3}{$S=1$}

    \ye{25.5}{17.5}{$\oplus$}
      \yt{27}{18}{1}
      \yd{27}{17}
      \yt{28}{18}{1}
      \yd{28}{17}
      \yt{29}{18}{1}
      \yt{29}{17}{2}
      \yt{30}{18}{1}
      \yt{30}{17}{2}
      \yt{31}{18}{2}
      \yd{31}{17}
      \yt{32}{18}{2}
      \yd{32}{17}
    \ye{29.5}{15.3}{$S=2$}
      \put(19,6){
      \ye{14.5}{11.5}{$\oplus$}
      \yt{16}{12}{1}
      \yd{16}{11}
      \yt{17}{12}{1}
      \yd{17}{11}
      \yt{18}{12}{1}
      \yd{18}{11}
      \yt{19}{12}{1}
      \yt{19}{11}{2}
      \yt{20}{12}{2}
      \yd{20}{11}
      \yt{21}{12}{2}
      \yd{21}{11}
      \yt{22}{12}{2}
      \yd{22}{11}

      \ye{19}{9.3}{$S=3$}
      \ye{23.5}{11.5}{$\oplus$}
      \yt{25}{12}{1}
      \yd{25}{11}
      \yt{26}{12}{1}
      \yd{26}{11}
      \yt{27}{12}{1}
      \yd{27}{11}
      \yt{28}{12}{1}
      \yd{28}{11}
      \yt{29}{12}{2}
      \yd{29}{11}
      \yt{30}{12}{2}
      \yd{30}{11}
      \yt{31}{12}{2}
      \yd{31}{11}
      \yt{32}{12}{2}
      \yd{32}{11}
      \ye{28.5}{9.3}{$S=4$}}
    \end{picture}
    \begin{picture}(52,5)(0,15) \linethickness{0.5pt}
       \yt{1}{18}{1}
       \yt{2}{18}{1}
       \yt{3}{18}{1}
       \yt{4}{18}{1}
       \yt{1}{17}{2}
       \yt{2}{17}{2}
       \yt{3}{17}{2}
       \yt{4}{17}{2}
     \ye{2.5}{15.3}{$S=0$}
     \ye{5.5}{18}{$\otimes$}
       \yt{7}{18}{3}
       \yt{8}{18}{3}
       \yt{9}{18}{3}
      \yt{10}{18}{3}
     \ye{8.5}{16.3}{$S=2$}

      \ye{12}{17.5}{$=$}
      \yt{14}{18}{1}
      \yt{14}{17}{2}
      \yt{15}{18}{1}
      \yt{15}{17}{2}
      \yt{16}{18}{1}
      \yt{16}{17}{2}
      \yt{17}{18}{1}
      \yt{17}{17}{2}
      \yt{18}{18}{3}
      \yt{19}{18}{3}
      \yt{20}{18}{3}
      \yt{21}{18}{3}
      \yd{18}{17}
      \yd{19}{17}
      \yd{20}{17}
      \yd{21}{17}
    \ye{17.5}{15.3}{$S=2$}
    \end{picture}
    \begin{picture}(52,5)(1,15) \linethickness{0.5pt}
      \put(1,0){\yt{0}{18}{1}
                \yt{1}{18}{1}
                \yt{1}{17}{2}
                \yt{2}{18}{1}
                \yt{2}{17}{2}
                \yt{3}{18}{1}
                \yt{3}{17}{2}
                \yt{4}{18}{2}
                \ye{2}{15.3}{$S=1$}}
      \ye{6.5}{18}{$\otimes$}
      \put(8,0){\yt{0}{18}{3}
                \yt{1}{18}{3}
                \yt{2}{18}{3}
                \yt{3}{18}{3}
              \ye{1.5}{16.3}{$S=2$}}
      \ye{13}{17.5}{$=$}
     \put(15,0){\yt{0}{18}{1}
                \yd{0}{17}
                \yt{1}{18}{1}
                \yt{1}{17}{2}
                \yt{2}{18}{1}
                \yt{2}{17}{2}
                \yt{3}{18}{1}
                \yt{3}{17}{2}
                \yt{4}{18}{2}
                \ye{4}{15.3}{$S=1$}}
     \put(19,0){\yt{0}{17}{3}
                \yt{1}{17}{3}
                \yd{1}{18}
                \yt{2}{18}{3}
                \yd{2}{17}
                \yt{3}{18}{3}   
                \yd{3}{17}}
     \ye{23.5}{17.5}{$\oplus$}
     \put(25,0){\yt{0}{18}{1}
                \yd{0}{17}
                \yt{1}{18}{1}
                \yt{1}{17}{2}
                \yt{2}{18}{1}
                \yt{2}{17}{2}
                \yt{3}{18}{1}
                \yt{3}{17}{2}
                \yt{4}{18}{2}
                \ye{3.5}{15.3}{$S=2$}}
     \put(29,0){\yt{0}{17}{3}
                \yt{1}{18}{3}
                \yd{1}{17}
                \yt{2}{18}{3}
                \yd{2}{17}
                \yt{3}{18}{3}   
                \yd{3}{17}}
     \put(20,5){
     \ye{13.5}{12.5}{$\oplus$}
     \put(15,-5){\yt{0}{18}{1}
                \yd{0}{17}
                \yt{1}{18}{1}
                \yt{1}{17}{2}
                \yt{2}{18}{1}
                \yt{2}{17}{2}
                \yt{3}{18}{1}
                \yt{3}{17}{2}
                \yt{4}{18}{2}
                \yd{4}{17}
                \ye{4}{15.3}{$S=3$}}
     \put(20,-5){\yt{0}{18}{3}
                \yd{0}{17}
                \yt{1}{18}{3}
                \yd{1}{17}
                \yt{2}{18}{3}
                \yd{2}{17}
                \yt{3}{18}{3}   
                \yd{3}{17}}}
    \end{picture}
    \begin{picture}(52,10)(2,10) \linethickness{0.5pt}
      \put(1,0){\yt{0}{18}{1}
                \yt{1}{18}{1}
                \yt{2}{18}{1}
                \yt{2}{17}{2}
                \yt{3}{18}{1}
                \yt{3}{17}{2}
                \yt{4}{18}{2}
                \yt{5}{18}{2}
                \ye{2.5}{15.3}{$S=2$}}
      \ye{7.5}{18}{$\otimes$}
      \put(9,0){\yt{0}{18}{3}
                \yt{1}{18}{3}
                \yt{2}{18}{3}
                \yt{3}{18}{3}
              \ye{1.5}{16.3}{$S=2$}}
      \ye{14}{17.5}{$=$}
     \put(16,0){\yt{0}{18}{1}
                \yd{0}{17}
                \yt{1}{18}{1}
                \yd{1}{17}
                \yt{2}{18}{1}
                \yt{2}{17}{2}
                \yt{3}{18}{1}
                \yt{3}{17}{2}
                \yt{4}{18}{2}
                \yt{5}{18}{2}
                \ye{3.5}{15.3}{$S=0$}}
     \put(20,0){\yt{0}{17}{3}
                \yt{1}{17}{3}
                \yt{2}{17}{3}
                \yd{2}{18}
                \yt{3}{17}{3}   
                \yd{3}{18}}
     \ye{24.5}{17.5}{$\oplus$}
     \put(26,0){\yt{0}{18}{1}
                \yd{0}{17}
                \yt{1}{18}{1}
                \yd{1}{17}
                \yt{2}{18}{1}
                \yt{2}{17}{2}
                \yt{3}{18}{1}
                \yt{3}{17}{2}
                \yt{4}{18}{2}
                \yt{5}{18}{2}
                \ye{3.5}{15.3}{$S=1$}}
     \put(30,0){\yt{0}{17}{3}
                \yt{1}{17}{3}
                \yt{2}{17}{3}
                \yd{2}{18}
                \yt{3}{18}{3}   
                \yd{3}{17}}
     \ye{34.5}{17.5}{$\oplus$}
     \put(36,0){\yt{0}{18}{1}
                \yd{0}{17}
                \yt{1}{18}{1}
                \yd{1}{17}
                \yt{2}{18}{1}
                \yt{2}{17}{2}
                \yt{3}{18}{1}
                \yt{3}{17}{2}
                \yt{4}{18}{2}
                \yt{5}{18}{2}
                \ye{3.5}{15.3}{$S=2$}}
     \put(40,0){\yt{0}{17}{3}
                \yt{1}{17}{3}
                \yt{2}{18}{3}
                \yd{2}{17}
                \yt{3}{18}{3}   
                \yd{3}{17}}
     \ye{44.5}{17.5}{$\oplus$}
     \put(46,0){\yt{0}{18}{1}
                \yd{0}{17}
                \yt{1}{18}{1}
                \yd{1}{17}
                \yt{2}{18}{1}
                \yt{2}{17}{2}
                \yt{3}{18}{1}
                \yt{3}{17}{2}
                \yt{4}{18}{2}
                \yd{4}{17}
                \yt{5}{18}{2}
                \ye{4}{15.3}{$S=3$}}
     \put(51,0){\yt{0}{17}{3}
                \yd{1}{17}
                \yt{1}{18}{3}
                \yt{2}{18}{3}
                \yd{2}{17}
                \yt{3}{18}{3}   
                \yd{3}{17}}

     \ye{14}{12.5}{$\oplus$}
     \put(16,-5){\yt{0}{18}{1}
                \yd{0}{17}
                \yt{1}{18}{1}
                \yd{1}{17}
                \yt{2}{18}{1}
                \yt{2}{17}{2}
                \yt{3}{18}{1}
                \yt{3}{17}{2}
                \yt{4}{18}{2}
                \yd{4}{17}
                \yt{5}{18}{2}
                \yd{5}{17}
                \ye{4.5}{15.3}{$S=4$}}
     \put(22,-5){\yt{0}{18}{3}
                \yd{0}{17}
                \yt{1}{18}{3}
                \yd{1}{17}
                \yt{2}{18}{3}
                \yd{2}{17}
                \yt{3}{18}{3}   
                \yd{3}{17}}
       \end{picture}
    \caption{Examples of products of extended tableaux for an $S=2$
      spin chain.}
    \label{fig:na2S2}
  \end{center}
\end{figure*}

In Figure \ref{fig:na2S2}, we illustrate the principle by writing
out a few terms in the expansion for an $S=2$ chain.
We now turn to the question what this construction implies for the
momentum spacings of the spinons.  It is very easy to see from
Figure \ref{fig:na2S2} that $b_1=1$ and $a_1$ is odd, and that
$b_L=2S$ and $a_L$ is even (odd) for $N$ even (odd).
Let us assume we have a spinon $i$
with momentum number $a_i$ and box number $b_i$.  If we take $S=3$, $a_i=3$,
and $b_i=2$, this spinon would be represented by a dot which shares a
column with the second box with number $3$ in it,
\begin{equation*}
  \setlength{\unitlength}{\ytlength}
  \begin{picture}(8,5)(0,0.5) \linethickness{0.5pt}
    \yt{1}{3}{\small 3}
    \yt{2}{3}{\small 3}
    \yt{3}{3}{\small 3}
    \yt{4}{3}{\small 3}
    \yt{5}{3}{\small 3}
    \yt{6}{3}{\small 3}
    \ye{7}{2.5}{.}
    \yd{2}{2}
    \ye{2.15}{0.8}{\small $b_i=2$}
  \end{picture}
\end{equation*}
For the box number $b_{i+1}$ of the next spinon, there are only two
possibilities:  

(i) $b_{i+1}=b_i+1$, which implies that $a_{i+1}-a_i$ is even.  The
spinons either sit in neighboring columns with $a_{i+1}=a_i$, or
contain an even number of spin $S$ representations (with $2S$ boxes
each) in between them.  For our example, the corresponding tableaux
are
\begin{equation*}
  \setlength{\unitlength}{\ytlength}
  \begin{picture}(28,5.5)(0,-0.8) \linethickness{0.5pt}
    \put( 0,0){\yt{1}{3}{\small 3}
      \yt{2}{3}{\small 3}
      \yd{2}{2}
      \ye{2.15}{0.8}{\small $b_i$}
      \yt{3}{3}{\small 3}
      \yd{3}{2}
      \ye{3.75}{0.7}{\small $b_{i+1}$}
      \ye{3.3}{-1}{\small $a_{i+1}=a_i$}
      \yt{4}{3}{\small 3}
      \yt{5}{3}{\small 3}
      \yt{6}{3}{\small 3}}
    \yd{2}{2}
    \ye{8.5}{2.5}{and}

    \put(10,0){\yt{1}{3}{\small 3}
      \yt{2}{3}{\small 3}
      \yd{2}{2}
      \ye{2.15}{0.8}{\small $b_i$}
      \yt{3}{3}{\small 3}
      \yt{4}{3}{\small 3}
      \yt{5}{3}{\small 3}
      \yt{6}{3}{\small 3}}
    \put(12,0){\yt{1}{2}{\small 4}
      \yt{2}{2}{\small 4}
      \yt{3}{2}{\small 4}
      \yt{4}{2}{\small 4}
      \yt{5}{3}{\small 4}
      \yt{6}{3}{\small 4}}

    \put(16,0){\yt{1}{2}{\small 5}
      \yt{2}{2}{\small 5}
      \yt{3}{3}{\small 5}
      \yd{3}{2}
      \ye{3.75}{0.7}{\small $b_{i+1}$}
      \ye{-.5}{-1}{\small $a_{i+1}=a_i+2$}
      \yt{4}{3}{\small 5}
      \yt{5}{3}{\small 5}
      \yt{6}{3}{\small 5}}
    \ye{24.5}{2.5}{and}
    \ye{27}{2}{$\ldots$}
  \end{picture}
\end{equation*}
This possibility produces the unidirectional, horizontal arrows in
Figure \ref{fig:biflow}.  If $b_i=2S$, this possibility does not
exist, and there are either no further spinons or $a_{i+1}-a_i$ has
to be odd.

(ii) $b_{i+1}=2S-b_i+1$, which implies that $a_{i+1}-a_i$ is odd.  For
our example, the tableaux are
\begin{equation*}
  \setlength{\unitlength}{\ytlength}
  \begin{picture}(32,5.5)(0,-0.8) \linethickness{0.5pt}
    \put( 0,0){\yt{1}{3}{\small 3}
      \yt{2}{3}{\small 3}
      \yd{2}{2}
      \ye{2.15}{0.8}{\small $b_i$}
      \yt{3}{3}{\small 3}
      \yt{4}{3}{\small 3}
      \yt{5}{3}{\small 3}
      \yt{6}{3}{\small 3}}
    \put(2,0){\yt{1}{2}{\small 4}
      \yt{2}{2}{\small 4}
      \yt{3}{2}{\small 4}
      \yt{4}{2}{\small 4}
      \yt{5}{3}{\small 4}
      \yd{5}{2}
      \ye{5.75}{0.7}{\small $b_{i+1}$}
      \ye{3.2}{-1}{\small $a_{i+1}=a_i+1$}
      \yt{6}{3}{\small 4}}
    \ye{10.5}{2.5}{and}

    \put(12,0){\yt{1}{3}{\small 3}
      \yt{2}{3}{\small 3}
      \yd{2}{2}
      \ye{2.15}{0.8}{\small $b_i$}
      \yt{3}{3}{\small 3}
      \yt{4}{3}{\small 3}
      \yt{5}{3}{\small 3}
      \yt{6}{3}{\small 3}}
    \put(14,0){\yt{1}{2}{\small 4}
      \yt{2}{2}{\small 4}
      \yt{3}{2}{\small 4}
      \yt{4}{2}{\small 4}
      \yt{5}{3}{\small 4}
      \yt{6}{3}{\small 4}}

    \put(18,0){\yt{1}{2}{\small 5}
      \yt{2}{2}{\small 5}
      \yt{3}{3}{\small 5}
      \yt{4}{3}{\small 5}
      \yt{5}{3}{\small 5}
      \yt{6}{3}{\small 5}}
    \put(20,0){\yt{1}{2}{\small 6}
      \yt{2}{2}{\small 6}
      \yt{3}{2}{\small 6}
      \yt{4}{2}{\small 6}
      \yt{5}{3}{\small 6}
      \yd{5}{2}
      \ye{5.75}{0.7}{\small $b_{i+1}$}
      \ye{-0.5}{-1}{\small $a_{i+1}=a_i+3$}
      \yt{6}{3}{\small 6}}
    \ye{28.5}{2.5}{$\ldots$}
  \end{picture}
\end{equation*}
This possibility produces the bidirectional, semicircle arrows in
Figure \ref{fig:biflow}.

This concludes the motivation of the flow diagram in Figure
\ref{fig:biflow}b.  As in Sections \ref{sec:hsyt} and \ref{sec:nana},
the single spinon momenta are given by
\begin{align}
  \label{eq:naasinglespinonmom}
  p_i=\frac{\pi}{N}\,\left(a_i-\frac{1}{2}\right).
\end{align}
This yields momentum spacings $p_{i+1}-p_i$ which can be either an
integer or an half-integer times $\frac{2\pi}{N}$.



\section{Correspondence with SU(2) level $k$ fusion rules}
\label{sec:co}
In this section, we establish a one-to-one correspondence between the
topological choices for the momentum spacings and the fusion rules of
spin \half spinons in the SU(2) level $k$ Wess--Zumino--Witten
model~\cite{wess-71plb95,witten84cmp455}.  The results we present are
completely consistent with, and in many ways complementary to, those
obtained by Bouwknegt, Ludwig, and
Schoutens~\cite{bouwknegt-94plb448,bouwknegt-95plb304,bouwknegt-96npb345,bouwknegt-99npb501}
based on the Yangian symmetry of the conformal field theory.

The ``deformed'' Lie algebra SU(2) level $k$ is in essence an SU(2)
spin algebra with the maximal spin restricted to $\frac{k}{2}$,
\begin{align}
  \label{eq:cok}
  j=0,\frac{1}{2},1,\frac{3}{2},\ldots ,\frac{k}{2}.
\end{align}
The fusion rules, and also the only possible rules consistent with
associativity, are given by
\begin{align}
  \label{eq:cojtimesj}
  j_1\otimes j_2= |j_1-j_2|&\oplus |j_1-j_2|+1 \oplus\ldots\nonumber\\[6pt]
  \ldots &\oplus\min\{j_1+j_2,k-j_1-j_2\}.
\end{align}
We will now investigate what these fusion rules imply for the spinons
of the models studied in the previous sections.

\begin{figure*}[htb!]
 \begin{center}
  \includegraphics[scale=1]{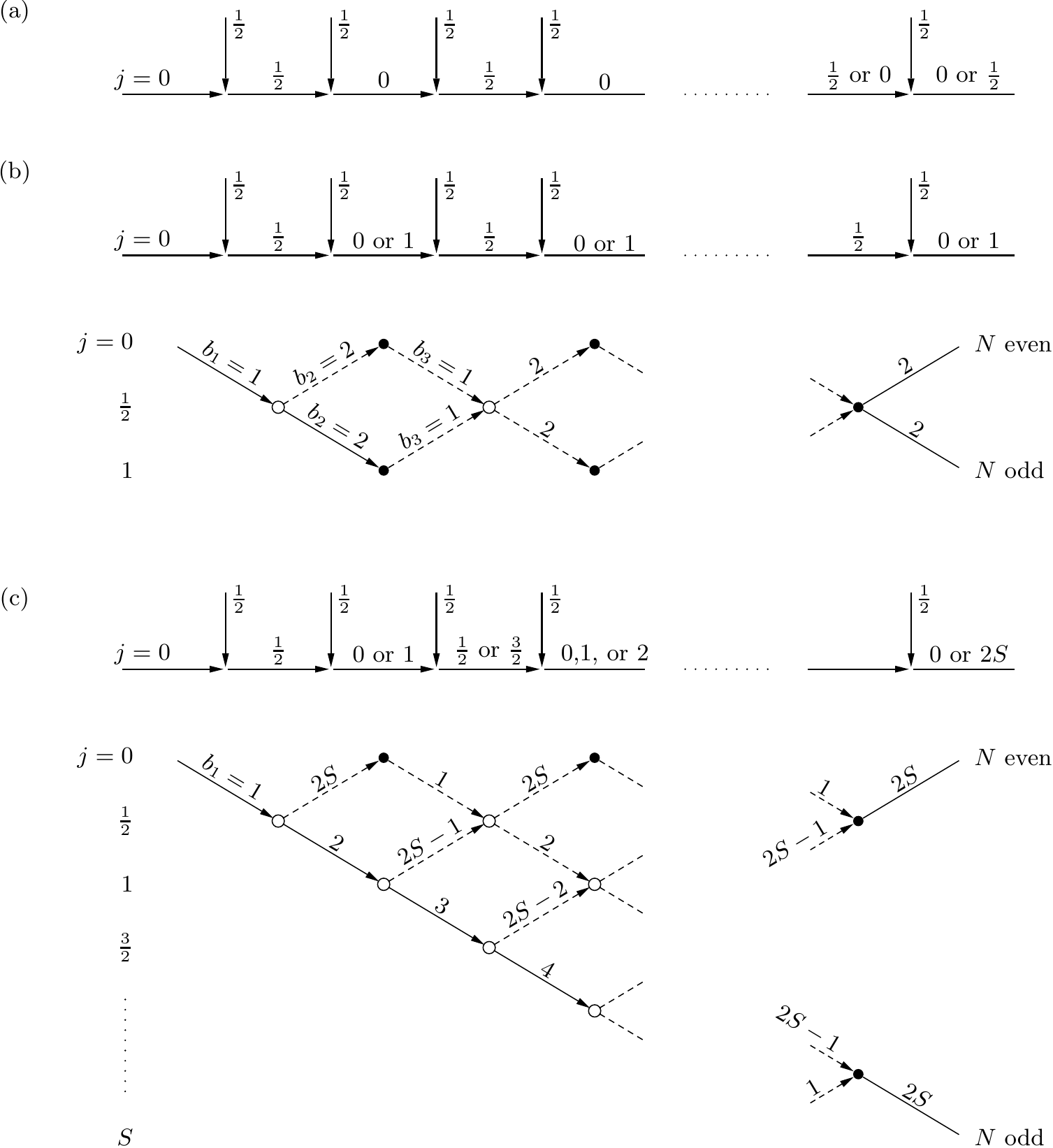}
  \end{center}
  \caption{Fusion trees and Bratteli diagrams for SU(2) level $k=2S$
    spinons: (a) In the Haldane--Shastry model ($k=1$), there is only
    one possible fusion trajectory for spinons with $j_i=\frac{1}{2}$,
    with $j$ on the links alternating between $0$ and $\frac{1}{2}$.
    The momentum spacings $a_{i+1}-a_i$ are always odd.  (b) For the
    critical $S=1$ model studied in Section \ref{sec:na}, the fusion
    rules \eqref{eq:cojtimesj} for Ising anyons ($k=2$) allow for a
    choice whenever we reach $j=\frac{1}{2}$, which is the case after
    every other spinon.  The momentum spacings $a_{i+1}-a_i$ are odd
    when we do not have a choice, and either even or odd when we have
    a choice.  (c) In the general $k=2S$ model studied in Section
    \ref{sec:naa}, there are many possible fusion trajectories, as
    illustrated by the dotted lines in the lower diagram.  We have a
    choice between two possiblities at all points where $j$ reaches
    neither $0$ nor $S$.  Continuation in the same direction, as
    illustrated for the first four spinons with a solid line,
    corresponds to following the unidirectional, horizontal arrows in
    Figure \ref{fig:biflow}b, which implies that $b_{i+1}=b_i+1$ and
    that $a_{i+1}-a_i$ is even.  Changing direction via a kink, on the
    other hand, corresponds to following one of the bidirectional,
    semicircle arrows in Figure \ref{fig:biflow}b, and implies that
    $b_{i+1}=2S+1-b_i$ and that $a_{i+1}-a_i$ is odd.}
  \label{fig:fusion}
\end{figure*}

We begin with the Haldane--Shastry chain, which is a microscopic
lattice realization of the SU(2) level $k=1$ Wess-Zumino-Witten
model.  For $k=1$, the relevant
fusion rules \eqref{eq:cojtimesj} are
\begin{align}
  \label{eq:cok1fusion}
  0\otimes\frac{1}{2}&=\frac{1}{2},\nonumber\\[5pt]
  \hspace{-1pt}\frac{1}{2}\hspace{1pt}\otimes\frac{1}{2}&=0,
\end{align}
\ie the fusion of two representations always yields a unique
representation.  The fusion diagram we obtain as we combine, starting
from $j=0$, spinons with $j=\frac{1}{2}$, is likewise unique, as
illustrated in Figure \ref{fig:fusion}a.  In the case of the
Haldane--Shastry model, the momentum spacings $a_{i+1}-a_i$ of the
spinons are always odd, in accordance with Abelian half-Fermi
statistics, a result we ultimately wish to relate to the fusion diagram.

We now turn to the $S=1$ model discussed in Section \ref {sec:na}.
The relevant fusion rules for $k=2$ are
\begin{align}
  \label{eq:k2fusion}
  0\otimes\frac{1}{2}&=\frac{1}{2},\nonumber\\[5pt]
  \hspace{-1pt}\frac{1}{2}\hspace{1pt}\otimes\frac{1}{2}&=0\oplus 1,
  \nonumber\\[5pt]
  1\otimes\frac{1}{2}&=\frac{1}{2},
\end{align}
\ie whenever we reach a representation $j=\frac{1}{2}$ on a horizontal
link in the fusion diagram, we have a choice between obtaining $j=0$
or $j=1$ when fusing it with the next spinon on the right, as
illustrated in Figure \ref{fig:fusion}b.  Since we reach
$j=\frac{1}{2}$ after every second spinon, the number of choices we
obtain is completely equivalent to the number we obtained via the
extended Young tableau formalism in Section \ref{sec:nana} (see also
Figure \ref{fig:biflow}b).  With this formalism, we further
established that the momentum spacing $a_{i+1}-a_i$ is odd whenever
fusing the previous spinon does not allow for a choice, and may be
either even or odd whenever the previous fusion provided us with a
choice.

We may hence tentatively assume that, whenever the direction of the
declining or ascending lines in the lower diagram in Figure
\ref{fig:fusion}b changes, the spacing $a_{i+1}-a_i$ between the two
spinons on both sides is odd (as for Abelian half-Fermi statistics),
and that whenever the direction of the line does not change, the
spacing $a_{i+1}-a_i$ is even (as for Abelian Bose or Fermi
statistics).

We will now confirm this assumption by elaborating the case of the
general spin $S$ model discussed in Section \ref{sec:naa}.  The
spinons of this model obey non-Abelian SU(2) level $k=2S$ statistics.
The relevant fusion rules are
\begin{align}
  \label{eq:k2sfusion}
  \hspace{1pt} 0\otimes\frac{1}{2}&=\frac{1}{2},\nonumber\\[5pt]
  \hspace{1pt} j\otimes\frac{1}{2}&=j-\frac{1}{2}\,\oplus\,
  j+\frac{1}{2},\quad\text{for}\ 0<j<S,\nonumber\\[5pt]
  S\otimes\frac{1}{2}&=S-\frac{1}{2},
\end{align}
\ie unless $j=0$ or $j=S$ on a horizontal link in the fusion diagram,
we have a choice of two possible representations when we fuse $j$ with
the next spinon, as illustrated in Figure \ref{fig:fusion}c.  Each
line in the lower diagram of this figure represents a spinon, and each
dot on its left represents the fusion of this spinon with the
representation $j$ from all the previous spinons.  The numbers on the
declining or ascending lines denote the box numbers $b_i$ of the
spinons, which are assigned according to the assumption we wish to
confirm.  As assigned, they are uniquely determined by the vertical
position and directions of the lines in the diagram.  Drawn with solid
lines on the left of the diagram is the simple sequence $b_1=1$,
$b_2=2$, $b_3=3$, $b_4=4$, which corresponds to following the
horizontal arrows in the flow diagram shown in Figure
\ref{fig:biflow}b.  This implies that the momentum spacings $a_4-a_3$,
$a_3-a_2$, and $a_2-a_1$ are even for this sequence.  The dotted lines
represent the alternatives.  Each time we change the direction of the
lines, say from declining to ascending as we add the second spinon,
the dot at the kink corresponds to one of the bidirectional,
semicircle arrows in Figure \ref{fig:biflow}b.  The kinks hence
represent even spacings $a_{i+1}-a_i$, as assumed above.

The diagram in Figure \ref{fig:fusion}c shows that there is a very
large number of possible trajectories.  Unless we have reached $j=0$
or $j=S$, both of which correspond to the last box $b_i=2S$
in Figure \ref{fig:biflow}b, we have the choice between continuing in
the direction of the line representing the previous spinon $i$ (which
implies $a_{i+1}-a_i$ even, as for Abelian Bose or Fermi statistics),
or changing the direction (which implies $a_{i+1}-a_i$ odd, as for
Abelian half-Fermi statistics).

With the last spinon, we must always reach $b_i=2S$.  If the number of
sites $N$ of the chain with PBCs is even, we must conclude with $j=0$.
It is not difficult to see from Figure \ref{fig:fusion}c that the
number of spinons must be even in this case.  If $N$ is odd, however,
we must conclude with $j=S$.  The number of spinons is given by $2S$
plus a non-negative, even integer.  These restrictions are completely
consistent with what we would expect from the projective construction
of the wave functions we discussed in Section \ref{sec:na:MSforSpinS}.
If $N$ is odd, we need at least one spinon in each of the $2S$
Haldane--Shastry chains we project together.  Further spinons can only
be created in pairs.

It is not really surprising that there is a one-to-one correspondence
between the fusion rules of SU(2) level $k=2S$ spinons with
$j=\frac{1}{2}$ and the rules for constructing the internal Hilbert
space for the SU(2) level $k=2S$ anyons we have derived in the
previous sections.  It is instructive, however, as this correspondence
provides us with the physical momentum spacings we obtain as we fuse
the $j=\frac{1}{2}$ representations of the individual spinons
according to the fusion diagram in Figure \ref{fig:fusion}c.


Since the notation of the momentum numbers $a_i$ is tied to the Young
tableau formalism introduced in Reference
~\onlinecite{greiter-07prl237202} and generalized to higher spin
models in the previous sections, it is appropriate to state the rules
for the momentum spacings in a generally more familiar language.  If
we fuse two SU(2) spinons in an SU(2) level $k=2S$ model (either one
of the microscopic models introduced by one of
us~\cite{Greiter11,nielsen-11jsmte11014} or the conformal field theory
of Wess, Zumino and Witten~\cite{wess-71plb95,witten84cmp455}, with
the spinon bases studied by Bouwknegt, Ludwig, and
Schoutens~\cite{bouwknegt-94plb448,bouwknegt-95plb304,bouwknegt-96npb345,bouwknegt-99npb501})
with $j=\frac{1}{2}$ each, and neighboring single spinon momenta $p_i$
and $p_{i+1}$, the spacing between these two momenta $p_{i+1}-p_i$ is
quantized according to Abelian half-Fermi statistics, if and only if
there is a kink between the spinons in the Bratteli diagram,
\psset{xunit=8pt,yunit=10pt,runit=14pt,linewidth=.4pt,doublesep=1.25\pslinewidth,arrowsize=2.6pt,arrowlength=2,arrowinset=0,arcangleA=85,arcangleB=85}
\begin{align}
  \label{eq:cosinglet}
\includegraphics{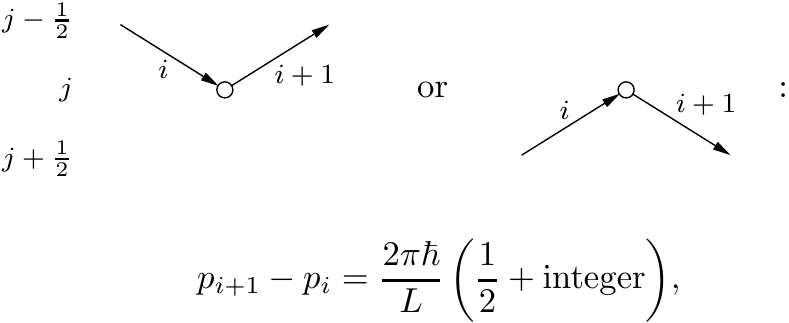}
\end{align}
where $L$ is the length of the chain with PBCs.  (``Neighboring'' here
means that there are no other spinons with momenta between $p_i$ and
$p_{i+1}$).  If, on the other hand, the two spinons form a straight
line in the Bratelli diagram, the spacing between the neighboring
single spinon momenta $p_i$ and $p_{i+1}$ is quantized according to
Bose or Fermi statistics (which are equivalent for our purposes here),
\begin{align}
  \label{eq:cotriplet}
  \includegraphics{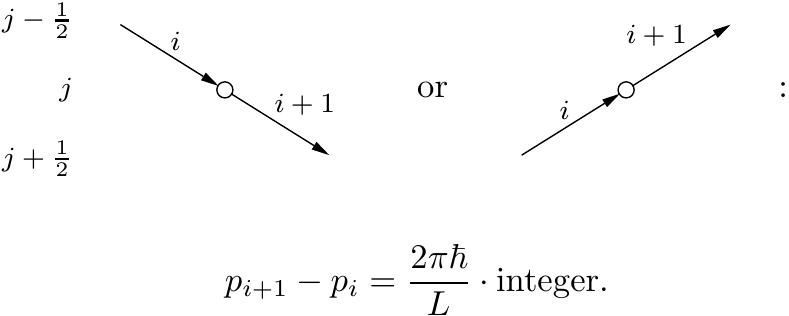}
\end{align}
These shifts are topological quantum numbers.  The spinon states with
different shifts become degenerate in the thermodynamic limit, but
since the states are topologically different, it is not possible to
connect them with local perturbation.  The situation here is
analogous to the Majorana fermion states in the vortex cores of the
quasiparticles of the Moore--Read state.


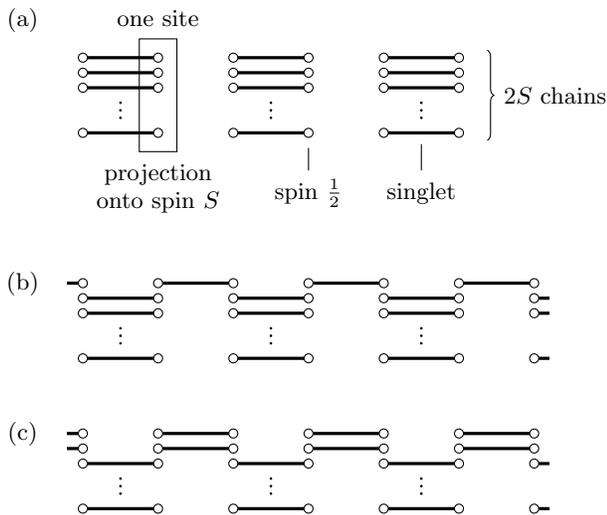
\begin{figure}[t]
  \centering
\begin{tikzpicture}[>=latex,scale=1]
\coordinate (Origin)   at (0,0);
\pgfmathsetmacro{\rdot}{0.06}
\def\cellcolor{black}
\newcommand{\dimer}[3]{
\begin{scope}[thin,shift={(#1,#2)} #3]] 
  \draw [fill=white, thin] (0,0) circle [radius=\rdot];
  \draw [fill=white, thin] (-1,0) circle [radius=\rdot];
  \draw[very thick](-\rdot,0)--(-1+\rdot,0);
\end{scope}
}
\newcommand{\dimerdot}[3]{
\begin{scope}[thin,shift={(#1,#2)} #3]] 
  \draw [fill=black, thin] (-.5,0) circle [radius=.01];
\end{scope}
}
\begin{scope}[shift={(0,0)}]
  \node at (-1.8,0.5){(a)};
  \begin{scope}  
    \clip(-1.2,-1.2) rectangle(4.2,0.2);
    \foreach \x in {0,2,4}{
      \foreach \y in {0,-.2,-.4,-1}{
        \dimer{\x}{\y}{}}
      \foreach \y in {-.6,-.7,-.8}{
        \dimerdot{\x}{\y}{}}
      }
  \end{scope}
  \draw[draw=\cellcolor](-.25,-1.25) rectangle (.25,.25);
  \node at (0,0.3)[\cellcolor,above]{\small one site};
  \node[align=center,\cellcolor,below] at (0,-1.3) {\small projection\\ onto spin $S$};
  \draw[decoration={brace,raise=5pt},decorate]
  (4.2,0.1) -- node[align=left,right=8pt] {\small $2S$ chains} (4.2,-1.1);
  \draw[thin] (2,-1.2)--(2,-1.5);
  \node at (2,-1.5)[below]{\small spin $\frac{1}{2}$};
  \draw[thin] (3.5,-1.15)--(3.5,-1.5);
  \node at (3.5,-1.55)[below]{\small singlet};
\end{scope}
\begin{scope}[shift={(0,-3)}]
  \node at (-1.8,0){(b)};
  \begin{scope}  
    \clip(-1.2,-1.2) rectangle(5.2,0.2);
    \foreach \x in {-1,1,3,5}{
      \foreach \y in {0}{
        \dimer{\x}{\y}{}}
      }
    \foreach \x in {0,2,4,6}{
      \foreach \y in {-.2,-.4,-1}{
        \dimer{\x}{\y}{}}
      \foreach \y in {-.6,-.7,-.8}{
        \dimerdot{\x}{\y}{}}
      }
  \end{scope}
\end{scope}
\begin{scope}[shift={(0,-5)}]
  \node at (-1.8,0){(c)};
  \begin{scope}  
    \clip(-1.2,-1.2) rectangle(5.2,0.2);
    \foreach \x in {-1,1,3,5}{
      \foreach \y in {0,-.2}{
        \dimer{\x}{\y}{}}
      }
    \foreach \x in {0,2,4,6}{
      \foreach \y in {-.4,-1}{
        \dimer{\x}{\y}{}}
      \foreach \y in {-.6,-.7,-.8}{
        \dimerdot{\x}{\y}{}}
      }
  \end{scope}
\end{scope}
\end{tikzpicture}
  \caption{(a)--(c) Illustration of three of the $2S+1$ possible dimer
    configurations for a spin $S$ chain.  The spin $S$ at each site of
    the chain is obtained by symmetric projection of the $k=2S$
    constituent spin $\frac{1}{2}$'s.
    The different dimerization patters are obtained from the reference
    pattern ($j=0$) shown in (a) by shifting $2j,\ j=0,\frac{1}{2},1,
    \frac{3}{2},\ldots ,S$ of the $2S$ dimerized
    constituent chains by one lattice spacing.  The critical SU(2)
    level $k$ spin chain can be thought of as the multi-critical point
    between all the possible dimer phases.}
  \label{fig:VBSgs}
\end{figure}

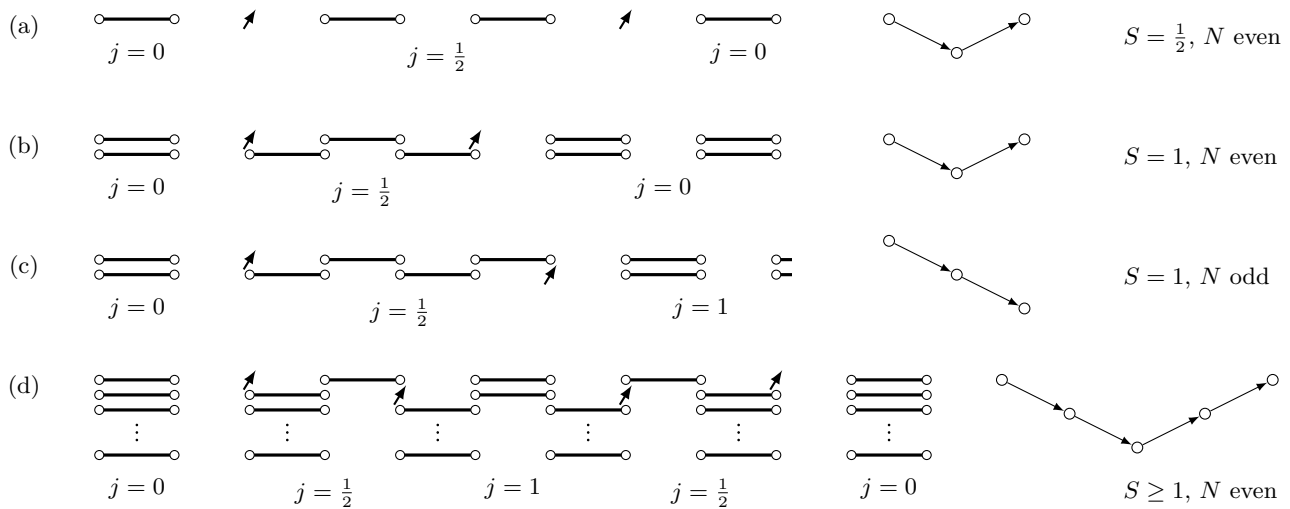
\begin{figure*}[t]
  \centering
\begin{tikzpicture}[>=latex,scale=1]
  \coordinate (Origin) at (0,0); 
  \pgfmathsetmacro{\rdot}{0.06}
  \pgfmathsetmacro{\chainsep}{.2} 
  \pgfmathsetmacro{\jlabelsep}{.2} 
  \def\cellcolor{black}
  \newcommand{\dimer}[3]{
\begin{scope}[thin,shift={(#1,#2)} #3]] 
  \draw [fill=white, thin] (0,0) circle [radius=\rdot];
  \draw [fill=white, thin] (-1,0) circle [radius=\rdot];
  \draw[very thick](-\rdot,0)--(-1+\rdot,0);
\end{scope}
}
\newcommand{\dimerdot}[3]{
\begin{scope}[thin,shift={(#1,#2)} #3]] 
  \draw [fill=black, thin] (-.5,0) circle [radius=.01];
\end{scope}
}
\newcommand{\spinon}[3]{
\begin{scope}[thin,shift={(#1,#2)} #3]] 
  \draw[->,thick](-.08,-.13)--(0.08,.13);
\end{scope}
}
\newcommand{\brattelinode}[3]{
  \node (#3) at (#1,#2) {};
  \draw [fill=white, thin] (#3) circle [radius=.08];
}
\newcommand{\jdwarrow}[2]{
  \draw[->] (#1) ++(.08,-.04)-- ++(0.84,-.42)(#2);
}
\newcommand{\juparrow}[2]{
  \draw[->] (#1) ++(.08,.04)-- ++(0.84,.42)(#2);
}

\begin{scope}[shift={(0,0)}]
  \node at (-2,-.1){(a)};
  \begin{scope}  
    \clip(-1.2,-1.2) rectangle(8.2,0.2);
    \foreach \x in {0,3,5,8}{
      \foreach \y in {0}{
        \dimer{\x}{\y}{}}
       };
    \foreach \x in {1,6}{
      \spinon{\x}{0}{}
      }
  \end{scope}
  \begin{scope}[shift={(0,-\jlabelsep)}]  
    \node at (-0.5,0)[below]{\small $j=0$};
    \node at (3.5,0)[below]{\small $j=\frac{1}{2}$};
    \node at (7.5,0)[below]{\small $j=0$};
  \end{scope}
  \begin{scope}[shift={(9.5,0)},scale=.9]
%
    \brattelinode{0}{0}{c1}
    \brattelinode{1}{-.5}{c2}
    \brattelinode{2}{0}{c3}
    \jdwarrow{c1}{c2}
    \juparrow{c2}{c3}
  \end{scope}
  \node at (12.5,-0.25)[right]{$S=\frac{1}{2}$, $N$ even};
\end{scope}
\begin{scope}[shift={(0,-1.6)}]
  \node at (-2,-.1){(b)};
  \begin{scope}  
    \clip(-1.2,-1.2) rectangle(8.2,0.2);
    \foreach \x in {0,3,6,8}{
      \dimer{\x}{0}{}
       };
    \foreach \x in {1,4}{
      \spinon{\x}{0}{}
      };
    \foreach \x in {0,2,4,6,8}{
      \dimer{\x}{-\chainsep}{}
       };
  \end{scope}
  \begin{scope}[shift={(0,-\chainsep -\jlabelsep)}]  
    \node at (-0.5,0)[below]{\small $j=0$};
    \node at (2.5,0)[below]{\small $j=\frac{1}{2}$};
    \node at (6.5,0)[below]{\small $j=0$};
  \end{scope}
  \begin{scope}[shift={(9.5,0)},scale=.9]
    \brattelinode{0}{0}{c1}
    \brattelinode{1}{-.5}{c2}
    \brattelinode{2}{0}{c3}
    \jdwarrow{c1}{c2}
    \juparrow{c2}{c3}
  \end{scope}
  \node at (12.5,-0.25)[right]{$S=1$, $N$ even};
\end{scope}
\begin{scope}[shift={(0,-3.2)}]
  \node at (-2,-.1){(c)};
  \begin{scope}  
    \clip(-1.2,-1.2) rectangle(8.2,0.2);
    \foreach \x in {0,3,5,7,9}{
      \dimer{\x}{0}{}
       };
    \foreach \x in {1}{
      \spinon{\x}{0}{}
      };
    \foreach \x in {0,2,4,7,9}{
      \dimer{\x}{-\chainsep}{}
       };
    \foreach \x in {5}{
      \spinon{\x}{-\chainsep}{}
      };
  \end{scope}
  \begin{scope}[shift={(0, -\chainsep -\jlabelsep)}]  
    \node at (-0.5,0)[below]{\small $j=0$};
    \node at (3,0)[below]{\small $j=\frac{1}{2}$};
    \node at (7,0)[below]{\small $j=1$};
  \end{scope}
  \begin{scope}[shift={(9.5,0.25)},scale=.9]
    \brattelinode{0}{0}{c1}
    \brattelinode{1}{-.5}{c2}
    \brattelinode{2}{-1}{c3}
    \jdwarrow{c1}{c2}
    \jdwarrow{c2}{c3}
  \end{scope}
  \node at (12.5,-0.25)[right]{$S=1$, $N$ odd};
\end{scope}
\begin{scope}[shift={(0,-4.8)}]
  \node at (-2,-.1){(d)};
  \begin{scope}  
    \clip(-1.2,-1.2) rectangle(10.2,0.2);
    \foreach \x in {0,3,5,7,10}{
      \dimer{\x}{0}{}
       };
    \foreach \x in {1,8}{
      \spinon{\x}{0}{}
      };
    \foreach \x in {0,2,5,8,10}{
      \dimer{\x}{-\chainsep}{}
       };
    \foreach \x in {3,6}{
      \spinon{\x}{-\chainsep}{}
      };
      \foreach \x in {0,2,4,6,8,10}{
        \dimer{\x}{-2*\chainsep}{}
        \foreach \y in {-3*\chainsep,-3.5*\chainsep,-4*\chainsep,}{
          \dimerdot{\x}{\y}{}}
        \dimer{\x}{-5*\chainsep}{}
      };
  \end{scope}
  \begin{scope}[shift={(0, -5*\chainsep -\jlabelsep)}]  
    \node at (-0.5,0)[below]{\small $j=0$};
    \node at (2,0)[below]{\small $j=\frac{1}{2}$};
    \node at (4.5,0)[below]{\small $j=1$};
    \node at (7,0)[below]{\small $j=\frac{1}{2}$};
    \node at (9.5,0)[below]{\small $j=0$};
  \end{scope}
  \begin{scope}[shift={(11,0)},scale=.9]
    \brattelinode{0}{0}{c1}
    \brattelinode{1}{-.5}{c2}
    \brattelinode{2}{-1}{c3}
    \brattelinode{3}{-.5}{c4}
    \brattelinode{4}{0}{c5}
    \jdwarrow{c1}{c2}
    \jdwarrow{c2}{c3}
    \juparrow{c3}{c4}
    \juparrow{c4}{c5}
  \end{scope}
  \node at (12.5,-1.5)[right]{$S\ge 1$, $N$ even};
\end{scope}
\end{tikzpicture}
  \caption{Examples of dimer configurations with two or four spinon
    excitations (domain walls) in (a) a spin \half chain, (b) and (c)
    a spin 1 chain, and (d) a spin chain with spin $S\ge 1$.  The
    chains are assumed to have periodic boundary conditions and even
    or odd number of sites $N$, as indicated.  Note that each domain
    wall shifts one constituent chain by one lattice spacing, and
    hence changes the configuration label $j$ by $\pm \frac{1}{2}$.
    It is hence trivial to read of the corresponding Bratteli
    diagrams, which are depicted to the right of each of the dimerized
    chains.  Note that even though no dimerization occurs in the
    multi-critical SU(2) level $k$ spin chain we discuss in Sections
    \ref{sec:na:MSforSpinS} and \ref{sec:co}, the Bratteli diagrams,
    and hence the topological properties (\ie momentum spacings) of
    the spinons are nonetheless identical to those of the spinons or
    domain walls in the dimerized chains.}
  \label{fig:VBS}
\end{figure*}

\section{Domain walls in valence bond phases}

One way to understand the SU(2) level $k$ Wess--Zumino--Witten model
for a spin chain with spin $S=\frac{k}{2}$ is as a description of the
multicritical point between the $k+1$ different ``dimerization'' or
valance bond solid (VBS) phases the spin chain can assume in the
ground state.  Three of the possible VBS configurations for a chain
with spin $S\ge 1$ are shown in Figure \ref{fig:VBSgs}.  Here,
the spin $S$ at each site is understood to be formed through the
completely symmetric projection of $k=2S$ spin $\frac{1}{2}$'s onto
one spin $S$.  Before projection, the constituent spin $\frac{1}{2}$'s
are arranged into dimers, indicated by the solid, horizontal lines,
which are just singlets formed by two spins at neighboring
sites on the same constituent chain.

In this framework, single spinon excitations are minimal domain walls
between the dimerized phases.  In Figure \ref{fig:VBS}, we illustrate
the correspondence between the spinons, the labels $j$ from Figure
\ref{fig:fusion}, and the Bratteli diagrams with a few examples.  Let
us first look at Figure \ref{fig:VBS}a.  The first spinons shifts the
dimer configurations from the initial one, to which we assign label
$j=0$, by one lattice spacing.  Following the terminology introduced
in Section \ref{sec:co}, we assign the label $j=\frac{1}{2}$ to the
chain where one constituent chain (and in this first example we have
only one constituent chain) has shifted by one lattice spacing.  If
the number of sites on the chain is even, we need an even number of
domain walls, and hence an even number of spinons in each constituent
chain to close the boundary.  This is in correspondence with the
Bratteli diagram shown to the right of the chain in Figure
\ref{fig:VBS}a.  Three of the many, possible spinon configurations for
a spin $S=1$ chain, and the corresponding Bratteli diagrams, are shown
in Figures \ref{fig:VBS}b, \ref{fig:VBS}c, and \ref{fig:VBS}d.  In
these pictures, the distance between spinons on the same constituent
chain in units of lattice spacings is an odd number, and an even
number for neighboring spinons on different constituent chains.
Making the connection with the results of Sections
\ref{sec:na:MSforSpinS} and \ref{sec:co}, we see that the momentum
spacing between neighboring spinons is according to
\eqref{eq:cosinglet} for spinons on the same constituent chain, and
according to \eqref{eq:cotriplet} for spinons on different constituent
chains.  Periodic chains with an odd number of sites require an odd
number of domain walls, and hence an odd number of spinons, on each
constituent chain.  The minimal number of spinons in this case is
$2S$, as illustrated for an $S=1$ chain in Figure \ref{fig:VBS}c. In
Figure \ref{fig:VBS}d, we give an example indicating how the
application of these rules proceeds to chains with general spin $S$.

Viewing spinons, which are by definitions excitations with spin \half
and no charge (as compared to electrons, which have spin \half and
charge $-1$, or spin flips, which carry spin 1 and no charge), as domain
walls in dimerized chains is intuitively rewarding.  It correctly
captures many subtle techical issues like the state counting, and
hence to a limited extent also the statistics, even though no
information regarding the physical manifestation of the fractional
statistics, the fractional shifts in the momentum spacings, can be extracted.  That
a direct correspondence between the dimer phases and the Bratteli
diagrams can be established, however, is remarkable since the chains
we describe are critical, and do not display any dimerization
patterns.  The reason is that we can view the critical, and for
$S>\frac{1}{2}$ multicritical, phases of the spin chains as the
critical points between all the possible dimer phases.  At
criticality, quantum fluctuations will have destroyed any possible
dimerization in the chains, but---and this is the crucial point---will
not have altered the topological properties of the spinon excitations
we can visualize as domain walls in dimerized chains.  The quantum
statistics of the spinons, is among these topological properties.  Of
course, the spinon excitations in the critical chains are not
localized in real space, and do not provide a context to consider
neighboring spinons in real space.  As we have argued in Section
\ref{sec:na:MSforSpinS}, the good and robust quantum numbers in the
critical chains are the topological shifts in the relative spacings
between neighboring spinons in momentum space, not in real space.

\section{Conclusion}

In this article, we have argued that in one dimension, non-Abelian,
and in particular SU(2) level $k$ statistics, manifests itself in
fractional, topological shifts in the spacings of neighboring
quasiparticle momenta $p_{i+1}-p_i$, and derived the general rules
which patterns for the shifts are allowed for each $k$.  For $k=2$,
the case of Ising anyons, we found that the state counting of the
internal Hilbert space associated with the non-Abelian statistics is
equivalent to that of Majorana fermion states attached to the spinons.
This led us to refer to the shifts as Majorana spacings. For this
case, the braiding of the vortices carrying Majorana fermions in the
two dimensional analog, the Pfaffian quantum Hall state, is also well
understood~\cite{ivanov01prl268}.  We consider it likely that we can
learn something about the open problem of the braiding properties of
SU(2) level $k>2$ vortices in Read--Rezayi states by exploring the
analogy to the one dimensional models we studied here.

The most important aspect of the emerging picture is the general
validity, which needs some clarification.  We have derived the
formalism for non-Abelian SU(2) level $k$ statistics using spin chains
with spin $S\ge\frac{1}{2}$ tuned to the multicritical point.
%
%
We found that the quantum statistics of the spinons in these systems
is encoded in the topological shifts in momentum spacings between
spinons with neighboring momenta, as detailed in Section
\ref{sec:na:MSforSpinS}.
The only case where the individual spinon momenta are known to be
perfectly good quantum numbers, however, is the Haldane--Shastry model
for spin $\frac{1}{2}$.  A natural question to ask, therefore, is what
happens in a general, critical or multi-critical, spin chain, when the
spinon momenta are not good quantum numbers.  Are our results, or is
our formalism, still applicable to the general case?  Fortunately, the
answer is yes.  The spacings between the adjacent momenta of the
particles carrying the Abelian or non-Abelian, fractional statistics
are given by the sum of the topological shifts derived above, 
and integer spacings which we have not specified.  In \eqref{eq:cosinglet}
and \eqref{eq:cotriplet}, the topological shifts are \half and 0, repectively.

The key point is that while there is no reason to assume that the
integers in \eqref{eq:cosinglet} and \eqref{eq:cotriplet}, are good
quantum numbers in a general system, the \emph{topological shifts
  are always good quantum numbers.}

The situation is comparable to fermions, bosons, and anyons in two
space dimensions.  There, the relative angular momentum between two
identical particles is an odd integer times $\hbar$ for fermions, an
even integer times $\hbar$ for bosons, and an even integer plus the
statistical parameter $\theta$ times $\hbar/\pi$ for anyons.  In a
many particle state in two-dimensions, like an electron liquid forming
a quantized Hall state or a liquid of quasiparticles with fractional
statistics forming a daughter fluid, the relative momentum between two
fermions or two anyons is never a good quantum number. The topological
shift, that is, the integer being odd or even for fermions or bosons,
respectively, and shifted by $\theta/\pi$ away from bosons for Abelian
anyons, however, is always a good quantum number.  This shift
specifies the (fractional) statistics.  Similarly, the momenta of the
Abelian and non-Abelian anyons in one space dimension do not have to
be good quantum numbers, while the topological shifts to the otherwise
integer momentum spacings we derived above, will always be good
quantum numbers.  These spacings specify the Abelian or non-Abelian
statistics, and are topological properties of the SU(2) level $k$ WZW
model.

An important, but by contrast unresolved, question is whether the
non-Abelian statistics of the one-dimensional models, if realized in a
multi-critical spin chain in the laboratory, could be used in quantum
computation or quantum cryptography.  Unfortunately, we have no
definite answers to report at the present time.

\begin{acknowledgments}
MG would like to thank Eddy Ardonne for discussions. MG and RT are
supported by the DFG through SFB 1170 ToCoTronics (project B04), and
further acknowledge financial support from the DFG through the
W\"urzburg-Dresden Cluster of Excellence on Complexity and Topology in
Quantum Matter -- \textit{ct.qmat} (EXC 2147, project-id
39085490). FDMH acknowledges funding from the Princeton Center for
Complex Materials, a MRSEC supported by NSF Grant DMR 1420541.
\end{acknowledgments}


\begin{thebibliography}{10}

\bibitem{wilczek90}
F. Wilczek, {\em Fractional statistics and anyon superconductivity} (World
  Scientific, Singapore, 1990).

\bibitem{stern10n187}
A. Stern, Nature {\bf 464},  187  (2010).

\bibitem{moore-91npb362}
G. Moore and N. Read, Nucl. Phys. B {\bf 360},  362  (1991).

\bibitem{greiter-92npb567}
M. Greiter, X.~G. Wen, and F. Wilczek, Nucl. Phys. B {\bf 374},  567  (1992).

\bibitem{read-00prb10267}
N. Read and D. Green, Phys. Rev. B {\bf 61},  10267  (2000).

\bibitem{ivanov01prl268}
D.~A. Ivanov, Phys. Rev. Lett. {\bf 86},  268  (2001).

\bibitem{stern-04prb205338}
A. Stern, F. von Oppen, and E. Mariani, Phys. Rev. B {\bf 70},  205338  (2004).

\bibitem{kitaev03ap2}
A.~Y. Kitaev, Ann. Phys. {\bf 303},  2  (2003).

\bibitem{nayak-08rmp1083}
C. Nayak, S.~H. Simon, A. Stern, M. Freedman, and S. Das~Sarma, Rev. Mod. Phys.
  {\bf 80},  1083  (2008).

\bibitem{read-99prb8084}
N. Read and E. Rezayi, Phys. Rev. B {\bf 59},  8084  (1999).

\bibitem{kitaev06ap2}
A. Kitaev, Ann.~of Phys. {\bf 321},  2  (2006).

\bibitem{yao-07prl247203}
H. Yao and S.~A. Kivelson, Phys. Rev. Lett. {\bf 99},  247203  (2007).

\bibitem{yao-11prl087205}
H. Yao and D.-H. Lee, Phys. Rev. Lett. {\bf 107},  087205  (2011).

\bibitem{greiter-09prl207203}
M. Greiter and R. Thomale, Phys. Rev. Lett. {\bf 102},  207203  (2009).

\bibitem{scharfenberger-11prb140404}
B. Scharfenberger, R. Thomale, and M. Greiter, Phys. Rev. B {\bf 84},  140404
  (2011).

\bibitem{greiter-14prb165125}
M. Greiter, D.~F. Schroeter, and R. Thomale, Phys. Rev. B {\bf 89},  165125
  (2014).

\bibitem{meng-15prb241106}
T. Meng, T. Neupert, M. Greiter, and R. Thomale, Phys. Rev. B {\bf 91},  241106
   (2015).

\bibitem{lecheminant-17prb140406}
P. Lecheminant and A.~M. Tsvelik, Phys. Rev. B {\bf 95},  140406  (2017).

\bibitem{glasser-15njp082001}
I. Glasser, J.~I. Cirac, G. Sierra, and A.~E.~B. Nielsen, New Journal of
  Physics {\bf 17},  082001  (2015).

\bibitem{liu-18prb195158}
Z.-X. Liu, H.-H. Tu, Y.-H. Wu, R.-Q. He, X.-J. Liu, Y. Zhou, and T.-K. Ng,
  Phys. Rev. B {\bf 97},  195158  (2018).

\bibitem{wildeboer-16prb045125}
J. Wildeboer and N.~E. Bonesteel, Phys. Rev. B {\bf 94},  045125  (2016).

\bibitem{Greiter11}
M. Greiter, {\em Mapping of Parent {H}amiltonians}, Vol.~244 of {\em Springer
  Tracts in Modern Physics} (Springer, Berlin/Heidelberg, 2011), published also
  as arXiv:1109.6104.

\bibitem{nielsen-11jsmte11014}
A.~E.~B. Nielsen, J.~I. Cirac, and G. Sierra, J. Stat. Mech.: Theory and
  Experiment {\bf 11},  P11014  (2011).

\bibitem{haldane88prl635}
F.~D.~M. Haldane, Phys. Rev. Lett. {\bf 60},  635  (1988).

\bibitem{shastry88prl639}
B.~S. Shastry, Phys. Rev. Lett. {\bf 60},  639  (1988).

\bibitem{wess-71plb95}
J. Wess and B. Zumino, Phys. Lett. {\bf 37B},  95  (1971).

\bibitem{witten84cmp455}
E. Witten, Commun. Math. Phys. {\bf 92},  455  (1984).

\bibitem{DiFrancescoMathieuSenechal97}
P. Di~Francesco, P. Mathieu, and D. S\'{e}n\'{e}chal, {\em Conformal Field
  Theory} (Springer, New York, 1997).

\bibitem{haldane91prl937}
F.~D.~M. Haldane, Phys. Rev. Lett. {\bf 67},  937  (1991).

\bibitem{greiter09prb064409}
M. Greiter, Phys. Rev. B {\bf 79},  064409  (2009).

\bibitem{greiter-07prl237202}
M. Greiter and D. Schuricht, Phys. Rev. Lett. {\bf 98},  237202  (2007).

\bibitem{scharfenberger-12jpa455202}
B. Scharfenberger and M. Greiter, Journal of Physics A: Mathematical and
  Theoretical {\bf 45},  455202  (2012).

\bibitem{thomale-12prb195149}
R. Thomale, S. Rachel, P. Schmitteckert, and M. Greiter, Phys. Rev. B {\bf 85},
   195149  (2012).

\bibitem{kopnin-91prb9667}
N.~B. Kopnin and M.~M. Salomaa, Phys. Rev. B {\bf 44},  9667  (1991).

\bibitem{bouwknegt-94plb448}
P. Bouwknegt, A.~W.~W. Ludwig, and K. Schoutens, Phys. Lett.~B {\bf 338},  448
  (1994).

\bibitem{bouwknegt-95plb304}
P. Bouwknegt, A.~W.~W. Ludwig, and K. Schoutens, Phys. Lett.~B {\bf 359},  304
  (1995).

\bibitem{bouwknegt-96npb345}
P. Bouwknegt and K. Schoutens, Nucl. Phys.~B {\bf 482},  345  (1996).

\bibitem{bouwknegt-99npb501}
P. Bouwknegt and K. Schoutens, Nucl. Phys.~B {\bf 547},  501  (1999).

\bibitem{affleck86npb409}
I. Affleck, Nucl. Phys. B {\bf 265},  409  (1986).

\bibitem{affleck-87prb5291}
I. Affleck and F.~D.~M. Haldane, Phys. Rev. B {\bf 36},  5291  (1987).

\bibitem{haldane91prl1529}
F.~D.~M. Haldane, Phys. Rev. Lett. {\bf 66},  1529  (1991).

\bibitem{shastry92prl164}
B.~S. Shastry, Phys. Rev. Lett. {\bf 69},  164  (1992).

\bibitem{haldane-92prl2021}
F.~D.~M. Haldane, Z.~N.~C. Ha, J.~C. Talstra, D. Bernard, and V. Pasquier,
  Phys. Rev. Lett. {\bf 69},  2021  (1992).

\bibitem{kawakami92prb1005}
N. Kawakami, Phys. Rev.~B {\bf 46},  1005  (1992).

\bibitem{talstra95}
J.~C. Talstra, Ph.D. thesis, Department of Physics, Princeton University, 1995.

\bibitem{bernevig-01prl3392}
B.~A. Bernevig, D. Giuliano, and R.~B. Laughlin, Phys. Rev. Lett. {\bf 86},
  3392  (2001).

\bibitem{greiter-06prl059701}
M. Greiter and D. Schuricht, Phys. Rev. Lett. {\bf 96},  059701  (2006).

\bibitem{gutzwiller63prl159}
M.~C. Gutzwiller, Phys. Rev. Lett. {\bf 10},  159  (1963).

\bibitem{kaplan-82prl889}
T.~A. Kaplan, P. Horsch, and P. Fulde, Phys. Rev. Lett. {\bf 49},  889  (1982).

\bibitem{metzner-87prl121}
W. Metzner and D. Vollhardt, Phys. Rev. Lett. {\bf 59},  121  (1987).

\bibitem{ha-93prb12459}
Z.~N.~C. Ha and F.~D.~M. Haldane, Phys. Rev.~B {\bf 47},  12459  (1993).

\bibitem{talstra-95jpa2369}
J.~C. Talstra and F.~D.~M. Haldane, J.~Phys.~A: Math. Gen. {\bf 28},  2369
  (1995).

\bibitem{ha-94prl2887}
Z.~N.~C. Ha and F.~D.~M. Haldane, Phys. Rev. Lett. {\bf 73},  2887  (1994),
  \emph{ibid.} \textbf{74}, E3501 (1995).

\bibitem{laughlin83prl1395}
R.~B. Laughlin, Phys. Rev. Lett. {\bf 50},  1395  (1983).

\bibitem{bernevig-01prb024425}
B.~A. Bernevig, D. Giuliano, and R.~B. Laughlin, Phys. Rev.~B {\bf 64},  024425
   (2001).

\bibitem{greiter-05prb224424}
M. Greiter and D. Schuricht, Phys. Rev.~B {\bf 71},  224424  (2005).

\bibitem{essler95prb13357}
F.~H.~L. E{\ss}ler, Phys. Rev.~B {\bf 51},  13357  (1995).

\bibitem{Hamermesh62}
M. Hamermesh, {\em Group Theory and its Application to Physical Problems}
  (Addison--Wesley, Reading, Mass., 1962).

\bibitem{InuiTanabeOnodera96}
T. Inui, Y. Tanabe, and Y. Onodera, {\em Group Theory and Its Applications in
  Physics} (Springer, Berlin, 1996).

\bibitem{greiter02jltp1029}
M. Greiter, J.~Low Temp. Phys. {\bf 126},  1029  (2002).

\bibitem{haldane83pl464}
F.~D.~M. Haldane, Phys. Lett. {\bf 93 A},  464  (1983).

\bibitem{haldane83prl1153}
F.~D.~M. Haldane, Phys. Rev. Lett. {\bf 50},  1153  (1983).

\bibitem{affleck90proc}
I. Affleck,  in {\em Fields, strings and critical phenomena}, Vol.~XLIX of {\em
  Les {H}ouches lectures}, edited by E. Br\'ezin and J. Zinn-Justin (Elsevier,
  Amsterdam, 1990).

\bibitem{Fradkin91}
E. Fradkin, {\em Field Theories of Condensed Matter Systems}, No.~82 in {\em
  Frontiers in Physics} (Addison Wesley, Redwood City, 1991).

\bibitem{affleck-87prl799}
I. Affleck, T. Kennedy, E.~H. Lieb, and H. Tasaki, Phys. Rev. Lett. {\bf 59},
  799  (1987).

\bibitem{affleck-88cmp477}
I. Affleck, T. Kennedy, E.~H. Lieb, and H. Tasaki, Commun. Math. Phys. {\bf
  115},  477  (1988).

\bibitem{greiter-07prb184441}
M. Greiter and S. Rachel, Phys. Rev. B {\bf 75},  184441  (2007).

\bibitem{greiter10np5}
M. Greiter, Nat. Phys. {\bf 6},  5  (2010).

\bibitem{michaud-13prb140404}
F. Michaud, S.~R. Manmana, and F. Mila, Phys. Rev. B {\bf 87},  140404  (2013).

\bibitem{schwinger65proc}
J. Schwinger,  in {\em Quantum Theory of Angular Momentum}, edited by L.
  Biedenharn and H. van Dam (Academic Press, New York, 1965).

\bibitem{Auerbach94}
A. Auerbach, {\em Interacting electrons and quantum magnetism} (Springer, New
  York, 1994).

\bibitem{frobenius-1882ram53}
G. Frobenius, J. Reine Angew. Math. {\bf 93},  53  (1882).

\bibitem{greiter-91prl3205}
M. Greiter, X.~G. Wen, and F. Wilczek, Phys. Rev. Lett. {\bf 66},  3205
  (1991).

\bibitem{nayak-96npb529}
C. Nayak and F. Wilczek, Nucl. Phys. B {\bf 479},  529  (1996).

\end{thebibliography}

\end{document}